\documentclass[aps,preprint,showpacs]{revtex4}
\usepackage[psamsfonts]{amsfonts,amssymb}
\usepackage{epsfig}
\usepackage{bm}

\begin{document}
\title{ A Hamiltonian system for interacting Benjamin-Feir resonances}

\author{T. Benzekri\email{benzekri@cpt.univ-mrs.fr}
\footnote{Corresponding Author: tel (+33) 491269522, fax (+33) 491269553}, C. Chandre\email{chandre@cpt.univ-mrs.fr}, R. Lima\email{lima@cpt.univ-mrs.fr}
and M. Vittot\email{vittot@cpt.univ-mrs.fr}}

\affiliation{Centre de Physique Th\'eorique, CNRS UMR 6207,
Luminy - case 907, F-13288 Marseille cedex 9, France}


\begin{abstract}
In this paper, we present a model describing the time evolution of two
dimensional surface waves in gravity and infinite depth. The model of
six interacting modes derives from the normal form of the system
describing the dynamics of surface waves and is governed by a Hamiltonian
 system of equations of cubic order in the amplitudes of the waves. We derive a
 Hamiltonian system with
 two degrees of freedom from this Hamiltonian using conserved quantities.
  The interactions are those of two coupled
Benjamin-Feir resonances. The temporal
evolution of the amplitude of the different modes is
described according to the parameters of the system. In particular,
we study the energy exchange produced by the modulations of the amplitudes of the modes.
The evolution of the modes reveals a chaotic
dynamics.
\end{abstract}

\maketitle

\section{Introduction}
\label{sec1}
In 1967, Benjamin and Feir~\cite{BeFe} studied the linear stability
of periodic solutions of permanent form for gravity surface waves
in infinite depth. They showed that wave trains are unstable to
modulation perturbation i.e that instabilities occur due to the
interaction of the carrier wave and two wave perturbations. This kind
of instability has been first found by Lighthill~\cite{Li} and
confirmed theoretically and experimentally by Benjamin and Feir. This
result was also confirmed by Zakharov~\cite{Za} using a Hamiltonian
formalism for the surface wave problem. This instability was
interpreted in terms of resonance conditions between the carrier wave
and the two side bands. Later in 1977, Lake {\it et al.}~
\cite{LaYeRuFe} observed in wave tank experiments the first
modulation and demodulation of the recurrence Fermi-Pasta-Ulam
phenomenon ~\cite{FPU}.\\ Many works using the Hamiltonian formalism
which was first derived by Zakharov, have been carried out to understand the
time evolution of the resonant interaction phenomenon ~\cite{Cr,CrGr1,DiKh}. We refer in
particular, for three dimensions, to Shemer and Stiassnie ~\cite{ShSt},
Stiassnie and Shemer~\cite{StSh},
 Shrira {\it et al.}~\cite{ShBaKh} and
Badulin {\it et al.}~\cite{BaShKhIo}. In two dimensions,
we refer to Benzekri {\it et al.}~\cite{BLV},
Caponi {\it et al.}
~\cite{CaSaYu} who use a seven modes model in order
to show a chaotic behavior. In order to study the long time evolution of
surface waves, Zufiria~\cite{Zu3} derived from the Zakharov water
waves equations a simple model of three interacting modes [Eq.(10) in Ref.~\cite{Zu3}]
which leads
to a chaotic behavior. However this model is not invariant by time
reversing symmetry.\\
Other works using the Hamiltonian formalism in fluid interfaces
 were carried out in Refs.~\cite{CrGr2,DiBr}.\\
In order to give a qualitative description of the spatio-temporal
evolution of the dynamical system described by a Hamiltonian water
wave system, we propose in our work a simple model using all the
symmetries of the initial equations of surface waves. In particular, we
use the time reversing symmetry to derive a system of Hamilton's
equations describing the dynamics of Benjamin-Feir resonant waves.
This symmetry plays an important role in the study of the system.
The evolution of the complex amplitudes of the modes is governed by a cubic dynamical system of
equations. The system is then a set of six ordinary differential
equations. We show the existence of invariants that reduces this system
to two degrees of freedom.

We use a geometrical approach to understand the energy exchange
mechanism among the modes of the wave. For some initial conditions,
we find that the Benjamin-Feir resonance leads to a continual periodic
energy transfer from one component of the wave to the others. We also find by numerical
simulations an irregular transfer of energy among the modes of the
wave. The energy transfer of this kind is related to
the existence of chaos in our system.

This paper is organized as follows: In Sec.~\ref{sec2}, we describe the
model which is an approximation of the Zakharov Hamiltonian water
waves system developed by Krazitskii \cite{Kra}. In this model,
the interactions require the coexistence of two resonating four-wave
interactions. The Hamiltonian of this model written in normal form is
of order four in the amplitudes of the wave.\\
 The
system has six degrees of freedom, then having six equations for the
complex amplitudes. We find four integrals of motion thanks to the
Manley-Rowe relations. Therefore, we reduce the model to a
Hamiltonian system with two degrees of freedom.

The knowledge of these invariants allows us to subdivide the phase
space in different regions and
 determine
analytical solutions like traveling and standing waves. In particular, we analyze how the system
behaves starting from different initial conditions. Section~\ref{sec3} is devoted to the study of
the dynamics of the system with only one resonance. The study of the chaotic dynamics
with the two interacting resonances is done in Sec.~\ref{sec4}.

\section{The model}
\label{sec2}

In an inviscid and incompressible fluid approximation, the equations
of the two-dimensional gravity waves on the surface of deep water
read as follows\cite{Za}:
\begin{eqnarray}
\label{141} && \Delta \varphi (x,y,t)=0, \qquad \mbox{  for }
-\infty <y<\eta (x,t), \\ &&g\eta (x,t)+\frac{\partial
\varphi}{\partial t}(x,y,t)+\frac{1}{2}| \nabla \varphi (x,y,t)|
^2=0,\quad \mbox{  for } y=\eta (x,t), \\ && \frac{\partial
\eta}{\partial t}(x,y,t)+\frac{\partial \varphi }{\partial x}
(x,y,t)\frac{\partial \eta}{\partial x}(x,y,t)-\frac{\partial
\varphi}{
\partial y}(x,y,t)=0,\quad \mbox{  for } y=\eta (x,t) ,
\\ && \lim_{y\rightarrow -\infty}\nabla \varphi (x,y,t)=0,
\end{eqnarray}
where the unknown variables are the surface shape of the wave \( \eta \),
and the stream function \( \varphi
\) and $g$ is the strength of the gravity.
 Zakharov~\cite{Za} showed that these
equations can be expressed in a Hamiltonian form:
\begin{equation}
\label{199}
\frac{da_{k}}{dt} = -i \frac{\partial{H}}{\partial{a^{*}_k}},
\qquad \mbox{  for } k  \in \mathbb{R} ,
\end{equation}
where the complex amplitudes $a_k$ are linear combinations of the
Fourier coefficients of $\eta(x,t)$ and of the free surface
velocity potential $\psi(x,t)$ defined by $$ \psi(x,t) =
\varphi(x,\eta(x,t),t). $$
If we restrict ourself to the spatially periodic waves then $k$ $\in$ $\mathbb{Z}$.
More precisely, if we define the
Fourier expansion of the wave $$
\eta(x,t)=\frac{1}{2\pi}\sum_{k=-\infty}^{+\infty}\eta_k e^{ikx},
$$ the relation between the coefficients $\eta_k$ and the complex
symplectic coordinates $a_k$ is $$ \eta_k(t)=\frac{g^{-1/4}}{\sqrt{2}} {\vert
k\vert} ^{1/4}\left( a_k+a_{-k}^{*}\right). $$ The
Hamiltonian is given by the total energy of the system when
expressed as function of the $a_k$ (see Ref.~\cite{Kra}).

Our model is obtained by truncating the general (infinite dimensional)
Hamiltonian derived by Zakharov in the form stated by Krazitszkii~\cite{Kra}
up to order four and retaining a minimal set of modes in order
to describe a Benjamin-Feir instability for gravity surface waves in
infinite depth. \\
In order to preserve time reversal symmetry, this minimal number of complex
modes is six, $\pm k_1$, $\pm
k_2$, $\pm k_3$ that leads to a set of two coupled resonances
verifying:
\begin{equation}
\label{626}
k_{1}+k_{2} = 2k_{3}, \quad \mbox{and} \quad \omega_{k_1}+\omega_{k_2}=2\omega_{k_3},
\end{equation}
together with the dispersion relation:
\begin{equation}
\label{1}
\omega_k= \sqrt{g |k|}, \quad \mbox{for} \quad k = \pm k_1, \pm k_2, \pm k_3.
\end{equation}

Dyachencho and Zakharov~\cite{DyZa} described the general families of such type of
resonances.

In our case this leads to only one family of resonances,
namely
$$
k_1=-a, \quad k_2=9a, \quad k_3=4a,  \quad \mbox{  for } a
\in {\mathbb Z}^{*} .\\
$$
For the simplicity of the exposition, the modes will be denoted by $a_j$ instead of
$a_{k_j}$ for $j=1,2,3$
and $a_{-j}$ instead of $a_{-k_j}$.\\
It is  clear from Eq.~(\ref{626}) that such resonances are built from central carriers,
$a_{3}$ and $a_{-3}$, and two equally distant side bands.\\
The expression of the Hamiltonian in the variables $a_j$ up to order 4 is
given in the Appendix.
Then following Ref.~\cite{Kra},  we perform a canonical transformation
from the variables $a_j$ to the variables $b_j$
such that the Hamiltonian is mapped into its normal form.
We rederive it  in the Appendix.\\
The Hamiltonian of the system becomes in the variables $b_j$
\begin{eqnarray}
H &=&\sum_{j=1,2,3}
\omega _{j}(| b_{j}| ^{2}+| b_{-j}| ^{2}) \nonumber
+\sum_{j=1,2,3} U_{j}( | b_{j}| ^{4}+ | b_{-j}|
^{4}-4| b_{j}| ^{2}| b_{-j}| ^{2}) \nonumber \\
&& +
V_{21}\;( | b_{2}| ^{2}| b_{1}| ^{2}+|
b_{-2}| ^{2}| b_{-1}| ^{2}) + V_{-21}\;( |
b_{2}| ^{2}| b_{-1}| ^{2}+| b_{-2}| ^{2}|
b_{1}| ^{2}) \nonumber \\
&& + V_{31}\;(- | b_{3}|
^{2}| b_{1}| ^{2}-| b_{-3}| ^{2}| b_{-1}|
^{2}) + V_{-31}\;( | b_{3}| ^{2}| b_{-1}|
^{2}+| b_{-3}| ^{2}| b_{1}| ^{2}) \nonumber \\
&& +
V_{23}\;( | b_{2}| ^{2}| b_{3}| ^{2}+|
b_{-2}| ^{2}| b_{-3}| ^{2}) + V_{-23}\;(- |
b_{2}| ^{2}| b_{-3}| ^{2}-| b_{-2}| ^{2}|
b_{3}| ^{2})\nonumber \\
&&+
 \frac{1}{2} W (b_{1}b_{2}b_{3}^{\ast
2}+b_{1}^{\ast}b_{2}^{\ast}b_{3}^{2}\;
+b_{-1}b_{-2}b_{-3}^{\ast 2}+b_{-1}^{\ast}b_{-2}^{\ast
}b_{-3}^{2}) , \label{504}
\end{eqnarray}
where the
coefficients $U_{j}$, $V_{ij}$ and $W$ are positive.
The values of all these coefficients are given in the Appendix .\\
The equations of motion are :
$$
\frac{db_{j}}{dt} = -i \frac{\partial{H}}{\partial{b^{*}_{j}}},
$$
for $j=-1,-2,-3,1,2,3$.

We notice that, in the infinite dimensional case, as showed by Dyachenko and Zakharov~\cite{DyZa}
and Craig and Worfolk~\cite{CrWo},
the canonical transformation that eliminates the cubic terms also cancels the fourth
order terms. In our case, starting from the finite dimensional truncation, the
Benjamin-Feir fourth order terms are instead present in the normal form. This
shows a difference between the infinite dimensional case and our finite
dimensional model. A similar Hamiltonian model for fluid interfaces has been derived in Ref.~\cite{CrGr2} where Benjamin-Feir resonances appear even in the infinite dimensional case due to the presence of the non-vanishing density of the upper fluid.\\

Using some scaling relations, it is possible to assume that
$\omega_1=1$, $\omega_2=3$ and $\omega_3=2$. Indeed, if we rescale
the variables $b_j$ by a factor $\sqrt{\lambda}$ for $\lambda$ $\in$ $\mathbb{R}^{*+}$
, i.e., if we replace the
Hamiltonian $H(b_j,b^{*}_j)$ by $\lambda H({ b_j}/\sqrt{\lambda},
{ b^{*}_j}/\sqrt{\lambda})$ which
is a transformation that preserves the equations of motion, and if
we rescale time by the same factor $\lambda$, i.e., we multiply
the Hamiltonian by $\lambda$, then it is easy to see that the
quartic part of the Hamiltonian is unchanged and the quadratic part is
multiplied by $\lambda$. By choosing $\lambda=1/\sqrt{g\vert
a\vert}$, it means that we can consider that the frequencies are
$\omega_1=1$, $\omega_2=3$ and $\omega_3=2$ without loss of generality. \\

As we shall see, the existence of several integrals of the motion in our system,
despite the fact that they are not sufficient to turn it into an integrable one,
allows to reduce the number of degrees of freedom from six to two and to  have
important information about the dynamics in phase space.\\
The equations of evolution of $\vert b_1\vert^2$, $\vert b_2\vert^2$ and $\vert b_3\vert^2$ are
\begin{eqnarray*}
&& \frac{d}{dt}\vert b_1\vert^2=i \frac{W}{2} (b_1b_2b^{*2}_3-b^{*}_1b^{*}_2b^{2}_3),\\
&& \frac{d}{dt}\vert b_2\vert^2=i \frac{W}{2} (b_1b_2b^{*2}_3-b^{*}_1b^{*}_2b^{2}_3),\\
&& \frac{d}{dt}\vert b_3\vert^2=-iW (b_1b_2b^{*2}_3-b^{*}_1b^{*}_2b^{2}_3).
\end{eqnarray*}
Therefore we see that the following two quantities are conserved
\begin{eqnarray}
\mathbb{I}_+ &=&2\vert b_1\vert^2+\vert b_3\vert^2 \; \label{517}, \\
\mathbb{J}_+ &=& \vert b_1\vert^2+ 3\vert b_2\vert^2+2 \vert b_3\vert^2. \label{518}
\end{eqnarray}
In what follows, we will also use a linear combination of these two conserved quantities
\begin{equation}
\mathbb{K}_+ =\vert b_1\vert^2 - \vert b_2\vert^2= \frac{1}{3}(2\mathbb{I}_+ - \mathbb{J}_+).\label{613}
\end{equation}
By defining $N_j \equiv \vert b_j\vert^2 $  for $j=1,2,3$, the constants of motion become:
\begin{eqnarray*}
\mathbb{I}_+ &=&2N_{1}+N_{3}, \\
\mathbb{J}_+ &=& N_{1}+3N_{2}+2 N_{3}.
\end{eqnarray*}
Similarly, we have two additional conserved quantities for the other resonance
\begin{eqnarray*}
\mathbb{I}_- &=&2N_{-1}+N_{-3}, \\
\mathbb{J}_- &=& N_{-1}+3N_{-2}+2 N_{-3},
\end{eqnarray*}
where $N_{-j} \equiv \vert b_{-j}\vert^2 $  for $j=1,2,3$.\\
It can be checked that the four integrals $\mathbb{I}_+$, $\mathbb{J}_+$, $\mathbb{I}_-$, $\mathbb{J}_-$
  are independent and in involution.
The key point to be noticed here, is the fact that each one of these first
integrals depends on the variables of one resonance alone even if, as it will be
seen, there is a dynamical interaction of the two sets of variables.\\

Let us now look at the structure of the invariant surfaces in phase space in view
of the previous integrals of motion. Even if they can not completely fix the
orbits of the system, they help to understand many properties of
the different dynamical regimes observed in this model.\\
In the space $\mathbb{E}_{+}=(N_{1}, N_{2}, N_{3})$
(respectively $\mathbb{E}_{-}=(N_{-1}, N_{-2}, N_{-3})$), $\mathbb{J}_{+}=const.$
  (resp. $\mathbb{J}_{-}=const.$)
defines a portion of a plane, that we represent in Fig.~\ref{Figure1}. On the other
hand, the intersection of this set with $\mathbb{I}_{+}=const.$
(resp. $\mathbb{I}_{-}=const.$) defines a family of segments $C_{+}(\mathbb{I}_+)$
(resp. $C_{-}(\mathbb{I}_-)$).\\

Let us consider the structure of $\mathbb{E}_+$ and let us fix $\mathbb{J}_{+} \ne 0$.
For $\mathbb{I}_{+}=0$, the set $C_+(0)$ reduces to one point $P$ whose coordinates are
$P=(0,\mathbb{J}_{+}/3,0)$.\\
For $0<\mathbb{I}_{+}<\frac{\mathbb{J}_{+}}{2}$, the intersection of the two
 invariant surfaces in $\mathbb{E}_{+}$, $C_{+}(\mathbb{I}_{+})$ is
a segment close to $P$.\\
When $\mathbb{I}_{+}=\mathbb{J}_{+}/2$,
$C_{+}(\mathbb{J}_{+}/2)$ is a segment with one extremity $T=
(0,0, \mathbb{J}_{+}/2)$.
We notice that on
this curve we have $N_1=N_2$.\\
For $\mathbb{J}_{+}/2<\mathbb{I}_{+}<2\mathbb{J}_{+}$, the set $C_{+}(\mathbb{I}_{+})$
is a segment close to $S=(\mathbb{J}_{+},0,0)$.\\
For $\mathbb{I}_{+}=2\mathbb{J}_{+}$, the intersection is the point $S$.\\
For $\mathbb{I}_{+}>2\mathbb{J}_{+}$ or for $\mathbb{I}_+ < 0$ the intersection is empty.

We notice that $\mathbb{E}_{+} \times \mathbb{E}_{-}$ needs to be supplemented
by additional variables like
the six angles $\Phi_{j}$ of the complex variables $b_j$ in order to build
the phase space. Therefore, since the projection of an orbit on
$\mathbb{E}_{+} \times \mathbb{E}_{-}$ alone is
only part of the information we must look at it as a partial, although useful,
information on the dynamics. This is why, in each case we shall supplement its
description with additional information in order to make a complete characterization
of the dynamics.

\section{Dynamics with  one resonance}
\label{sec3}
To get insight into the full dynamics, we begin with simple and integrable
situations which will allow us to
understand the general case with  two resonances. We start exploring the phase space by
choosing initial conditions leading  to simple situations for which we can
exhibit different kinds of  shapes of the wave. A complete description of phase
space is given by a numerical analysis.\\ It follows from
the conserved quantities,
that an orbit corresponding to an
initial condition with all null coordinates in one of the two resonance variables $N_{j}$
will keep this property for any time.
For instance, we assume that $\mathbb{J}_-=0$. This implies that $N_{-1}=N_{-2}=N_{-3}=0$
for all time. Thus, we first study the dynamics with only one resonance which is an integrable
case since Hamiltonian~(\ref{100}) has three degrees of freedom and three conserved quantities
(e.g. $H$, $\mathbb{I}_+$, $\mathbb{K}_+$).
Hamiltonian~(\ref{504}) describing the dynamics in the variables
$(b_1, b_2, b_3)$ becomes :

\begin{eqnarray}
H &=&
| b_{1}| ^{2}+ 3 | b_{2}| ^{2}+ 2| b_{3}| ^{2}
+ U_{1} | b_{1}| ^{4} + U_{2} | b_{2}| ^{4} + U_{3} | b_{3}| ^{4}
 \nonumber \\
&& +
V_{21} | b_{2}| ^{2}| b_{1}| ^{2}
- V_{31} | b_{3}|^{2}| b_{1}| ^{2}
 +
V_{23} | b_{2}| ^{2}| b_{3}| ^{2}
\nonumber \\
&&+\frac{1}{2} W (b_{1}b_{2}b_{3}^{\ast
2}+b_{1}^{\ast}b_{2}^{\ast}b_{3}^{2}).  \label{100}
\end{eqnarray}

In what follows, we describe the dynamics on the sets $C_+(\mathbb{I}_+)$ described in the
previous section (see Fig.~\ref{Figure1}). We  start with the simplest cases which
are the three points $P$, $T$ and $S$ and the segment $(PS)$ (one of the boundaries of
the triangle  $\mathbb{E}_+$)
for which we can exhibit
analytical expressions for the shape of the wave $\eta(x,t)$.\\

\subsection{Dynamics of orbits of $P$,$T$,$S$}
\label{sec31}
First we consider  the orbits corresponding
to the point $P$ of $\mathbb{E}_+$
(where $\mathbb{I}_+=0$).
The equations of motion  read
\begin{eqnarray*}
&& N_1 \equiv | b_{1}| ^{2}=0, \\
&& N_3 \equiv | b_{3}| ^{2}=0,\\
&& N_{2} \equiv | b_{2}| ^{2}=\frac{\mathbb{J}_{+}}{3},\\
&& i\frac{d b_{2}}{dt}=(3+2U_2\vert b_2\vert^2)b_2.
\end{eqnarray*}
The solution in the variables $b_j$ is then
\begin{eqnarray}
\label{600} b_{2}(t)&=&\sqrt{\frac{\mathbb{J}_{+}}{3}}
\exp\biggl(-i(3+2U_2\frac{\mathbb{J}_{+}}{3})t-i\Phi\biggr), \nonumber \\
b_{j}(t)&=&0, \qquad \mbox{for } j=1,3.
\end{eqnarray}
Therefore for any $\mathbb{J}_{+}>0$ these orbits are periodic in time.
Since we have only one resonance, the original variables $a_j$ are equal to $b_j$
(see Remark 1 of the Appendix).\\
Thus the expression of the shape of the wave is:
\begin{equation}
 \eta (x,t) = \frac{\sqrt{3}g^{-1/4}}{2\pi\sqrt{2}}
 \left( b_2(t) \exp(9ix)+
  b_2^{*}(t) \exp(-9ix)\right),
\end{equation}

which is also
\begin{equation}
\label{564} \eta (x,t) = \eta_{0}   \cos(9(x-c_2t)+\phi),
\end{equation}
i.e. traveling waves with velocity
\begin{equation}
c_2= \frac{1}{3}+\frac{4 \pi^2}{9} \sqrt{g} U_2 {\eta_{0}}^2.
\end{equation}

A plot of the shape of this wave is represented on Fig.~\ref{Figure2}.\\
Similarly for $S$, we obtain traveling waves running backwards
$\eta(x,t)=\eta_{0} \cos(x+c_1 t +\phi)$ with velocity
$$
c_1 =1+ {4 \pi^2} \sqrt{g} U_1 {\eta_{0}}^2.
$$
For $T$, we obtain traveling waves running forwards $\eta(x,t)=\eta_{0} \cos(4(x-c_3t)+\phi)$
with velocity
$$
c_3= \frac{1}{2}+ {\pi^2} \sqrt{g} U_3 {\eta_{0}}^2.
$$

 A linear stability of the points $P$,$S$ and $T$ shows that these fixed points
of the dynamics are elliptic. It means that nearby trajectories will mimic
the behavior of the dynamics of these points.
In particular, the surface waves near these points are time quasiperiodic waves close
to traveling waves.\\

\subsection{Dynamics on a point on the segment $(PS)$}
\label{sec32}
We consider the point which is at the intersection of $C_{+}(\mathbb{I}_{+})$ and the
segment $(PS)$,
for a value of $\mathbb{I}_+$ $\in$ $\left[0,2 \mathbb{J}_+\right] $.
This point is such that $b_3(0)=0$ and $ d{b_3(0)}/dt=0$. From the equation for $b_3(t)$,
 we deduce that $b_3(t)=0$
for any time.
From the two invariants $ \mathbb{I}_+$ and  $ \mathbb{J}_+$,
we have that $N_1$ and $N_2$ are constant and equal to $N_1(t)=\mathbb{I}_+/2$
and $N_2(t)=(2\mathbb{J}_+-\mathbb{I}_+)/6$.
The equations of motion become:
\begin{eqnarray*}
\frac{d{b_1}}{dt}&=&-i\left( 1+ (U_1-\frac{V_{21}}{6}) \mathbb{I}_+
+ \frac{V_{21}}{3}\mathbb{J}_+\right) b_1, \\ \nonumber
\frac{d{b_2}}{dt}&=&-i\left( 3+ (-\frac{U_2}{3}+\frac{V_{21}}{2}) \mathbb{I}_+
+ \frac{2}{3}V_{21} \mathbb{J}_+\right) b_2.
\end{eqnarray*}
We deduce the expression of  ${b_1}$ and  ${b_2}$:
\begin{eqnarray*}
b_1(t)&=& \sqrt{\frac{\mathbb{I}_+}{2}}
\exp\left( -i\left[ 1+ (U_1-\frac{V_{21}}{6}) \mathbb{I}_+
+ \frac{V_{21}}{3}\mathbb{J}_+\right] t+\alpha_1\right),
 \\ \nonumber
 &=& \sqrt{\frac{\mathbb{I}_+}{2}} \exp\left[{-i(c_1t+\alpha_1)}\right],\\ \nonumber
b_2(t)&=&\sqrt{(2\mathbb{J}_+-\mathbb{I}_+)/6}
\exp \left( -i\left[ 3+ (-\frac{U_2}{3}+\frac{V_{21}}{2}) \mathbb{I}_+
+\frac{2}{3}V_{21} \mathbb{J}_+ \right] t+\alpha_2\right),
\\ \nonumber
&=& \sqrt{\frac{2\mathbb{J}_+-\mathbb{I}_+}{6}} \exp\left[{-i(9c_2t+\alpha_2)}\right],
\end{eqnarray*}
where $c_1=1+ (U_1-{V_{21}}/6) \mathbb{I}_+ + {V_{21}}/3\mathbb{J}_+$
and $c_2=\left[ 3+ (-{U_2}/3+V_{21}) \mathbb{I}_+ +{2}/3V_{21} \mathbb{J}_+
\right]/9$.\\
By taking into account  Remark 1 of
 the Appendix, we have $a_j=b_j$ for $j=1,2,3$.
 The expression of the shape of the wave is:
\begin{eqnarray*}
\eta(x,t)&=&\frac{g^{-1/4}}{{2}\pi }
 \left( \sqrt{\mathbb{I}_+} \cos(x+c_1t+\alpha_1) \right. \\ \nonumber
&& +\left. \sqrt{2\mathbb{J}_+-\mathbb{I}_+} \cos(9(x-c_2 t)+\alpha_2) \right),
\end{eqnarray*}
which is the sum of two traveling waves with two different velocities. \\
We notice that on $P$ (respectively on  $S$), the amplitude of the traveling wave with
velocity $c_1$ (respectively $c_2$)
vanishes. Therefore the wave associated with $P$ (respectively $S$) becomes the traveling wave
with velocity $c_2$ (respectively $c_1$) obtained in Sec.~\ref{sec3}.\\

\subsection{Dynamics of orbits on $C_{+}(\mathbb{I}_{+})$}
\label{sec33}
Next, we extend the dynamics with one resonance, by studying
the dynamics  on the full sets $C_{+}(\mathbb{I}_{+})$ for $\mathbb{I}_{+}$ $\in$
$\left[0, 2 \mathbb{J}_{+}\right]$ .
We determine numerically  the profiles of the waves which are similar to those
obtained previously namely stationary,
traveling, time periodic and quasiperiodic waves. This study gives also
all the information about the exchanges of energy which occur between the modes.\\
In order to study the orbits in this case, since we have a Hamiltonian with three degrees
of freedom and three constants of motion, we want to perform a canonical
change of variables
reducing  the dimension of the system to one degree of freedom. Furthermore,
 we would like to change canonically the coordinates into action-angle variables
 $(N_j,\Phi_j)$ such that $b_j=\sqrt{N_j} e^{-i\Phi_j}$.\\
First we notice that if $N_3(0)=0$, then $N_3(t)=0$ at any time t (see Sec.~\ref{sec32}). In what
follows, we always exclude this  point which is at the intersection of $(PS)$ and
$C_{+}(\mathbb{I}_{+})$.\\
In the case where $\mathbb{I}_{+}= \mathbb{J}_+/2$, we notice that since
$\mathbb{K}_{+}=N_1-N_2$ is a conserved quantity (equal to zero in this case), if $N_1$
vanishes at some time, then $N_2$ also vanishes. This dynamics occurs at the  point $T$
 (see Sec.~\ref{sec31}). Except this particular case, there are two interesting cases where
$\mathbb{I}_{+}> \mathbb{J}_+/2$ and $\mathbb{I}_{+}< \mathbb{J}_+/2$.
If  $\mathbb{I}_{+}< \mathbb{J}_+/2$, only $N_1$  vanishes since
$N_2 \geq (\mathbb{J}_+-2\mathbb{I}_{+})/3 >0 $. If  $\mathbb{I}_{+}> \mathbb{J}_+/2$,
only $N_2$  vanishes since $N_1 \geq (2 \mathbb{I}_+-2\mathbb{J}_{+})/3 >0$.\\

We remark that $N_1$ is such that $\rm{max}(0,\mathbb{K}_{+}) \leq N_1 \leq \mathbb{I}_{+}/2$,
so the accessible phase space in the  $(p_1,q_1)=(\rm{Re}(b_{1}),\rm{Im}(b_{1}))$-plane
is
a circle for $\mathbb{K}_{+}<0$ and an annulus for $\mathbb{K}_{+}>0$.\\
In order to avoid the two cases $N_1=0$ or $N_2=0$ for which the angle $\phi_1$ or
$\phi_2$ is not defined
in the change of variables, we first show that
there is only one trajectory $\mathcal{T}_1$ such that $N_1(t)=0$,
i.e which passes by zero in the $(p_1,q_1)$-
plane. \\
First we notice that Hamilton's equations associated with Hamiltonian (\ref{100})
are invariant by the two transformations:
\begin{eqnarray*}
\hat{b}_1&=&{b}_1,  \nonumber \\
\hat{b}_2&=&{b}_2  e^{-i\phi}, \nonumber \\
\hat{b}_3&=& {b}_3 e^{-i\phi/2},
\end{eqnarray*}
and
\begin{eqnarray*}
\hat{b}_1&=&{b}_1 e^{-i\phi}, \nonumber \\
\hat{b}_2&=&{b}_2,   \nonumber \\
\hat{b}_3&=& {b}_3  e^{-i\phi/2},
\end{eqnarray*}
where $\phi$ is constant.
Using these two invariants, one can show that there is basically one trajectory $\mathcal{T}_1$
which passes by zero in the $(p_1,q_1)$-plane.
The closed trajectory $\mathcal{T}_1$ divides the phase space
in the $(p_1,q_1)$-plane
into two parts: the inside and the outside of $\mathcal{T}_1$.\\
Similarly, for $\mathbb{K}_{+}>0$, there is only one trajectory
$\mathcal{T}_2$ which passes by zero in $(p_2,q_2)$-plane.\\
We notice that  $N_2=0$ corresponds to  $N_1=\mathbb{K}_{+}$ and so a  trajectory
which passes by zero in the $(p_2,q_2)$-plane, is a  trajectory which passes by
the point $N_1=\mathbb{K}_{+}$ in the $(p_1,q_1)$-plane, which is the inside
border of the accessible region.\\
The trajectory $\mathcal{T}_2$ divides the annulus into two parts:
the inside and the outside of $\mathcal{T}_2$. Thus in any of these two parts, $N_1$ and
$N_2$ are strictly positive.\\
In the regions where all actions $N_1$, $N_2$ and $N_3$ are strictly positive,
we perform a canonical transformation that expresses Hamiltonian~(\ref{100})
into action-angle variables~:
\begin{eqnarray}
\label{1000}
b_{j}=\sqrt{N_j}e^{-i\Phi_j}.
 \end{eqnarray}
This transformation is canonical i.e. :
$$
\sum_{j=1,2,3} i db_{j} \wedge db_{j}^*=\sum_{j=1,2,3} dN_j\wedge d\Phi_j.
$$
The actions $N_1$, $N_2$, $N_3$ are respectively
conjugate to the angles $\Phi_1$, $\Phi_2$, $\Phi_3$. Hamiltonian (\ref{100}) becomes
\begin{eqnarray}
H&=&
N_1+3N_2+2N_3 + U_{1}N_1^2 +U_{2}N_2^2 +U_{3}N_3^2\nonumber \\
&& + V_{21}N_1 N_2 - V_{31}N_1N_3 + V_{23}N_2N_3\nonumber \\
&&+W\sqrt{N_1N_2}N_3\cos (\Phi_1+\Phi_2-2\Phi_3). \label{102}
\end{eqnarray}
Hamilton's equations for Hamiltonian~(\ref{102}) are given by
\begin{eqnarray*}
\frac{d\Phi_j}{dt}=\frac{\partial H}{\partial N_j},\\
\frac{dN_j}{dt}=-\frac{\partial H}{\partial \Phi_j},
\end{eqnarray*}
for $j=1,2,3$.\\

Using the invariants $\mathbb{I}_+$ and  $\mathbb{J}_+$, we now proceed
to the reduction of the dimension of
the system to one degree of freedom.\\
We perform the following canonical transformation from $(N_1,N_2,N_3,\Phi_1,
\Phi_2,\Phi_3)$
to \\ $(\tilde{N_1},\mathbb{K}_+,\mathbb{I}_+,\theta_1,\theta_2,\theta_3)$:

\begin{eqnarray}
\tilde{N_1}&=&N_1, \nonumber  \\
\mathbb{K}_+&=&N_1-N_2, \nonumber  \\
\mathbb{I}_+&=&2N_1+N_3, \nonumber  \\
\theta_1&=&\Phi_1+\Phi_2-2\Phi_3, \nonumber  \\
\theta_2&=&-\Phi_2, \nonumber    \\
\theta_3&=&\Phi_3.\label{105}
\end{eqnarray}
 The actions $\tilde{N_1}$,
$\mathbb{K}_+$ and $\mathbb{I}_+$ are now respectively conjugate to angles
$\theta_1$, $\theta_2$ and $\theta_3$. In what follows, we use $N_1$
instead of $\tilde{N}_1$ for simplicity. Hamiltonian~(\ref{102}) becomes

\begin{eqnarray}
H
=\Omega_{0}+\Omega_{1}N_{1}+\Omega_{2}N_{1}^{2}
+W\sqrt{N_{1}}\sqrt{N_{1}-\mathbb{K}_{+}}(\mathbb{I}_{+}-2N_{1})\cos
\theta _{1} , \label{103}
\end{eqnarray}
where the functions $\Omega_0$ and $\Omega_1$  only depend
on the actions $\mathbb{I}_{+}$, $\mathbb{K}_{+}$,
and $\Omega_2$  is constant~:
\begin{eqnarray}
\Omega_{0}&=&-3\mathbb{K}_{+}+2\mathbb{I}_{+}+U_{3}\mathbb{I}_{+}^{2}
+U_{2}\mathbb{K}_{+}^{2}
-V_{23}\mathbb{I}_{+}\mathbb{K}_{+}, \nonumber\\
\Omega_{1}
&=&-2U_{2}\mathbb{K}_{+}-4U_{3}\mathbb{I}_{+}-V_{21}\mathbb{K}_{+}
-V_{31}\mathbb{I}_{+}  +V_{23}(\mathbb{I}_{+}+2\mathbb{K}_{+}), \nonumber \\
\Omega_{2}
&=&U_{1}+U_{2}+4U_{3}+V_{21}+2V_{31}-2V_{23}.
\end{eqnarray}
Numerically  $\Omega_{2}$  is given by
$\Omega_{2}=-3.9642$.\\
We notice that Hamiltonian~(\ref{103}) does not depend on the angles $\theta_2$, and
$\theta_3$ which is equivalent to saying that $\mathbb{I}_{+}$ and  $\mathbb{K}_{+}$ are
invariant (constants of motion).\\
In order to visualize the dynamics, we perform numerical integration of the
equations of motion associated with Hamiltonian~(\ref{103}).\\
A plot of some trajectories for  $\mathbb{J}_{+}=1$  and different values of
$\mathbb{I}_{+}$ in the accessible
region of the phase space (grey region) is given in Fig.~\ref{Figure3}.
 In these plots, we have
transformed from action-angle $(N_{1},\theta_{1})$ to cartesian coordinates
$P_{1}=\sqrt{N_1}\cos{\theta_{1}},Q_{1}=\sqrt{N_1}\sin{\theta_{1}}$.
In these last variables, the accessible region is, for $\mathbb{K}_{+} <0$, i.e.
$\mathbb{I}_{+}<\mathbb{J}_{+}/2$,
the set of  points in the  interior of
the  circle $N_1=\mathbb{I}_{+}/2$ and such that these points are
inside  $\mathcal{T}_1$ or outside
$\mathcal{T}_1$ (where the canonical transformation from the original variables
$b_j$ to action-angle variables ($N_j$, $\theta_j$) is well defined).
 For $\mathbb{K}_{+} >0$, i.e. $\mathbb{I}_{+}>\mathbb{J}_{+}/2$, the accessible region is
the  annulus $\mathbb{K}_{+} < N_1< \mathbb{I}_{+}/2$
and such that these points are inside  $\mathcal{T}_2$ or outside
$\mathcal{T}_2$.\\
We clearly see that these plots are characterized by elliptic, hyperbolic points and the
associated
 separatrices. Below, we discuss the location of these points with respect
to the parameter $\mathbb{I}_{+}$.\\
The fixed points are obtained when  $ \dot{N}_{1} =0$ and  $\dot{\theta} _{1} =0$.
As depicted in
 Fig.~\ref{Figure1}, there is three families of fixed points.
The first family, denoted  $e(\mathbb{I}_{+})$, is located at $\theta_{1}=0$
 and is elliptic. Its position $N_1$ is solution
of the equation:
\begin{eqnarray}
\label{epf}
 \Omega_{1}+2 \Omega_{2} N_{1}
 + \frac{W}{2}
 \left[  \left( \sqrt{\frac{N_{1}-\mathbb{K}_{+}}{N_{1}}}
 +\sqrt{\frac{N_{1}}{N_{1}-\mathbb{K}_{+}} }\right)  (\mathbb{I}_{+}-2N_{1})
 -4  \sqrt{N_{1}(N_{1}-\mathbb{K}_{+})} \right] =0. \nonumber \\
\end{eqnarray}
which is the equation of $\dot{\theta}_{1} =0$ for  ${\theta}_{1}=0$.\\
There exist  two others families of fixed points located at $\theta_{1}=\pi$,
one  elliptic point denoted
$e_1(\mathbb{I}_{+})$ and the other  hyperbolic point denoted
$h_1(\mathbb{I}_{+})$.\\
The location $N_1$ is given by the equation:
\begin{eqnarray}
\label{epf1}
 \Omega_{1}+2 \Omega_{2} N_{1}
 -\frac{W}{2}
 \left[  \left( \sqrt{\frac{N_{1}-\mathbb{K}_{+}}{N_{1}}}
 +\sqrt{\frac{N_{1}}{N_{1}-\mathbb{K}_{+}} }\right)  (\mathbb{I}_{+}-2N_{1})
 -4  \sqrt{N_{1}(N_{1}-\mathbb{K}_{+})} \right] =0. \nonumber \\
\end{eqnarray}
which is the equation of $\dot{\theta} _{1} =0$ for  $\theta _{1}=\pi$.\\
The location of all these fixed points as a function of  $ \mathbb{I}_{+} $
is depicted on  Fig.~\ref{Figure1}.\\
We see on the  different plots of Fig.~\ref{Figure3}, that when $ \mathbb{I}_{+} $ is  close to zero,
there  is no fixed points. When we increase $ \mathbb{I}_{+} $ there  is
only one elliptic fixed point
$e(\mathbb{I}_{+})$( Fig.~\ref{Figure3}(a) ) which appears near zero.
 As  the value of  $ \mathbb{I}_{+} $
increases, there are three fixed points:
besides the previous elliptic
fixed point $e(\mathbb{I}_{+})$, there is a bifurcation (fold, which is  structurally stable)
where two new fixed points appear,
 an elliptic point $e_1(\mathbb{I}_{+})$
and a hyperbolic one $h_1(\mathbb{I}_{+})$(Fig.~\ref{Figure3}(b)).

We have a homoclinic connection for the hyperbolic fixed point. When the value of
$ \mathbb{I}_{+} $  still increases and remains lower than $ {\mathbb{J}_{+}}/{2} $, the
elliptic fixed point $e(\mathbb{I}_{+})$ moves to the right and  the hyperbolic
one $h_1(\mathbb{I}_{+})$
moves to the left, both  going towards the border of the phase space. At the same time,
the elliptic fixed point $e_1(\mathbb{I}_{+})$ moves to the right going to the
center of  the phase space so approaching $N_1=0$.
When $ \mathbb{I}_{+} $ is close to $ {\mathbb{J}_{+}}/{2} $,
the elliptic fixed point $e_1(\mathbb{I}_{+})$ remains near the border $N_1=0$ but it
is not apparent in the figures and the
hyperbolic one $h_1(\mathbb{I}_{+})$  remains near the border $N_1=\mathbb{I}_{+}/{2}$.
(Fig.~\ref{Figure3}(c)).\\
For $ \mathbb{I}_{+}> {\mathbb{J}_{+}}/{2}$
the hyperbolic point  $h_1(\mathbb{I}_{+})$
remains at the border  $N_1=\mathbb{I}_{+}/{2}$  and
the elliptic point $e_1(\mathbb{I}_{+})$ is in the border of the annulus where
$ \mathbb{K}_{+}>0 $.
At the same time
the region in which $N_1$ evolves decreases ( Fig.~\ref{Figure3}(d) and Fig.~\ref{Figure3}(e)).
For $ \mathbb{I}_{+}$ near $2 {\mathbb{J}_{+}}$, the phase space merges to the point $S$.\\
We notice that for $ \mathbb{I}_{+}= {\mathbb{J}_{+}}/{2}$, the resonance occupies a larger
 region of  phase space than the other cases. This is in agreement with
 Fig.~\ref{Figure1}.
All these fixed points correspond to periodic variations in variables
$N_{1}, N_{2}, N_{3} $   and $ \Phi_1, \Phi_2, \Phi_3 $. Indeed, due to the invariants
given by Eqs.(\ref{517})-(\ref{518}), if $ N_{1}$ is fixed, $ N_{2}$ and $N_{3} $ remain fixed and  the angles
$ \Phi_1, \Phi_2, \Phi_3 $  evolve linearly in time.
For example in the case  $\mathbb{I}_{+}=\mathbb{J}_{+}/2 $  (
i.e. $\mathbb{K}_{+}=0$), we  give explicitly the dynamics corresponding
to $e(\mathbb{J}_{+}/2)$.
 The equations of motion  are
\begin{eqnarray}
\frac{dN_{1}}{dt} &=&W N_{1}(\mathbb{I}_{+}-2N_{1})\sin \theta _{1},
\nonumber \\ \frac{d\theta _{1}}{dt} &=& \Omega_{1}+ W\mathbb{I}_{+}\cos \theta _{1}
+2N_1(\Omega_{2} - 2
W\cos \theta_{1}). \label{618}
\end{eqnarray}
For $\theta_1=0$ and
\begin{equation} \label{619} N_{1}= \frac{\Omega_{1}+W\mathbb{I}_{+}}{-2\Omega_{2}+4 W},
\end{equation}
we get an elliptic fixed point with $ N_1 \leq \mathbb{I}_{+}/2$ for all
$\mathbb{I}_{+}=\mathbb{J}_{+}/2 $. This means that this fixed point exists for all
 admissible values of  the parameter $\mathbb{J}_{+}$ and for $\mathbb{I}_{+}=\mathbb{J}_{+}/2 $ .\\
We deduce the expressions of the linear motion of the angles~:

\begin{eqnarray} \label{620}
\Phi_1(t)&=&\left[  1-V_{31}\mathbb{I} _{+}+\frac{1}{2}W\mathbb{I}
_{+}+(2U_1+V_{21}-2V_{31}-W)N_1\right]t+\Phi_1(0),\nonumber  \\
&=&c_1t+\alpha_1 ,\nonumber
\\ \Phi_2(t)&=&\left[3+V_{23}\mathbb{I} _{+}+\frac{1}{2}W\mathbb{I}
_{+}+(V_{21}+2U_2-2V_{23}-W)N_1\right]t+\Phi_2(0),\nonumber \\
&=&9c_2 t+\alpha_2 ,\nonumber
\\ \Phi_3(t)&=&\left[2+2U_3\mathbb{I} _{+}
+(-V_{31}+V_{23}-4U_3+W)N_1\right]t+\Phi_3(0),\nonumber\\
 &=&4c_3 t+\alpha_3 .\nonumber
\end{eqnarray}
This linear dynamics gives the following shape of the wave:
\begin{eqnarray} \label{622}
\eta(x,t)&=&\frac{g^{-1/4}}{\pi\sqrt{2}}
 \left( \sqrt{N_1} \cos{(x+c_1t+\alpha_1)}\right. \nonumber \\
&&+\sqrt{3}
\sqrt{N_1} \cos{[9(x-c_2 t)+\alpha_2]}
\nonumber\\
&&+\left.  \sqrt{2}
\sqrt{\mathbb{I} _{+}-2N_1}\cos{[4(x-c_3t)+\alpha_3]} \right) ,
\end{eqnarray}
which is a sum of three traveling waves with different velocities.\\
For this case, we have plotted on Fig.~\ref{Figure4} the elevation of the surface $\eta(x,t)$.
We see that $\eta$ is quasiperiodic in time. \\

\subsection{Exchanges of energy on $C_{+}(\mathbb{I}_{+})$} 
\label{sec34}
We will now see how  the exchanges of energy occur among the three modes
for different values of $\mathbb{I}_{+}$.\\
Since all the orbits of $C_{+}(\mathbb{I}_{+})$ are closed, the function $N_1(t)$
is periodic. From the two invariants $\mathbb{I}_{+}$ and $\mathbb{J}_{+}$,
the two other actions $N_2(t)$ and $N_3(t)$ are periodic with the same period as $N_1(t)$.
Therefore, all the exchanges which occur in the system are periodic in time.
\begin{itemize}
\item We consider $C_{+}(\mathbb{I}_{+})$ with
$\frac{\mathbb{J}_{+}}{2}<\mathbb{I}_{+}<2\mathbb{J}_{+}$. Let us
first consider the trajectories with initial conditions close to
$S$, i.e., such that $N_{2}(0)$ and $N_3(0)$ are of order
$\varepsilon<<1$. If we set
$\mathbb{I}_+=2 \mathbb{J}_+-\varepsilon$, we get from
Eqs.~(\ref{517})--(\ref{518})
\begin{eqnarray*}
&&\frac{\mathbb{I}_+}{2}-\frac{\varepsilon}{6} \leq N_{1}(t)
\leq \frac{\mathbb{I}_+}{2}, \\
&&  0 \leq N_{2}(t)\leq \frac{\varepsilon}{6},\\
&& 0 \leq N_{3}(t)\leq \frac{\varepsilon}{3},
\end{eqnarray*}
for all $t$. \\ In fact, this behavior is valid for all values of $\mathbb{I}_{+}$.
So, if we have initially $N_1(0) >> N_2(0)$ and $N_3(0)$,
then by looking at the invariant $\mathbb{I}_{+}$, we see that for $N_3(0)$ close to zero,
$N_1(t)$
decreases and at the same time $N_3(t)$ increases. The invariant
$2N_{2}+N_{3}\approx 2 N_{2}(0)$ shows that $N_2(t)$ also
decreases  and $N_3(t)$ does not grow  more than $2 N_{2}(0)$ which is small
 as we can  see
 in Fig.~\ref{Figure5}, where we have plotted the time evolution of the
 amplitudes $N_{1}$, $N_{2}$ and $N_{3}$ for the initial condition
 $(N_1,\theta_1)=(\frac{\mathbb{I}_{+}}{2},\pi)$.\\
Therefore, when almost all the energy is initially carried out by
the lower frequency mode $N_{1}$, the oscillations of the amplitudes are very
small and there is no significant energy exchange between the modes.\\
By looking directly at Fig.~\ref{Figure1}, we also see that, for such values
of $\mathbb{I}_{+}$, only the
modes $N_{2}$ and $N_{3}$ may exchange their energy while $N_{1}$ may only
sustain small energy fluctuations. \\
In the same way, for an orbit corresponding to $C_{+}(\mathbb{I}_{+})$
 with $0<{\mathbb{I}_{+}}<\mathbb{J}_{+}/2$,
there is no significant energy exchanges for initial conditions close to $P$,
and in any case only $N_{1}$ and $N_{3}$ enable some energy exchange among the
modes.

\item In order to obtain large energy exchanges between the three modes, we have to put
as much energy in the side bands as in the central carrier: the contribution of
energy to the side bands is equal to $\omega_1N_1+\omega_2N_2=N_1+3N_2$.
The maximum energy exchange is obtained when  $N_1+3N_2=2N_3=\mathbb{J}_{+}/2$. This happens
in the zone where $N_{1}$ and  $N_{2}$ are equally
important namely close to $C_{+}(\mathbb{J}_{+}/2)$.\\
In this case, using the form of the three
integrals of the motion and assuming
$\mathbb{I}_{+}=\mathbb{J}_{+}/2+\varepsilon$, we get:
\begin{equation}
0 \leq N_{3}(t) \leq \frac{\mathbb{J}_{+}}{2}+\varepsilon,
\quad \frac{2\varepsilon}{3}
\leq N_{1}(t) \leq \frac{\mathbb{J}_{+}}{4}+
\frac{\varepsilon}{2}.
\end{equation}
Then if we assume  that at $t=0$ $N_3(0) >> N_1(0)$ and $N_2(0)$ such that
the most energy is stored in the highest frequency mode, from
$2N_{1}+N_{3}\approx N_{3}(0)$
and $N_{1}(t)-N_{2}(t)=N_{1}(0)-N_{2}(0)$ we see that $N_3(t)$ decreases and at the
same time $N_{1}(t)$ and $N_{2}(t)$ increase.
So nothing prohibits that energy will be completely transferred from the
highest frequency mode into the low frequency modes.
What we observe is a periodic modulation of the wave amplitudes. We
see that most of the initial energy  carried out by the central frequency mode
of the wave, $N_{3}$, is transferred to the side bands and the magnitude of the
amplitude oscillations is very large. This is the usual Benjamin-Feir instability:
the carrier and the two side-bands evolve on a periodic cycle of modulation and
demodulation known as Fermi-Pasta-Ulam oscillations.
 In  Fig.~\ref{Figure6}
we see how the energy
exchange is done among the amplitudes $N_{1}$, $N_{2}$ and $N_{3}$ for for initial
$(N_1,\theta_1)=({\mathbb{I}_{+}}/{8},0)$.\\For this initial condition,
we have plotted in Fig.~\ref{Figure7}, the shape of the wave $\eta(x,t)$
for this initial condition.
The periodic modulation and demodulation of the three modes give a shape of the wave
which is a modulated traveling wave.

\item We can also see exchanges of energy for $\mathbb{I}_{+}=\mathbb{J}_{+}/2$  i.e. on
the segment $ C_{+}(\mathbb{J}_{+}/2)$ and for   initial conditions so that
 $N_{3}(0)$ is close to zero and
 $N_{1}$, $N_{2}$ lies on  $ C_{+}(\mathbb{J}_{+}/2)$
so that $N_{1}(0)=N_{2}(0)\approx\mathbb{J}_{+}/4$.
We recall that if $N_1(0)=N_2(0)$ then $N_1(t)=N_2(t)$ and $\mathbb{K}_{+}=0$.\\
For this case, from Eq.(\ref{517}), we conclude that $N_1(t)$ and $N_2(t)$
decrease and $N_3(t)$
grows until it reaches its maximum which is   $\mathbb{J}_{+}/4$.
As shown in Fig.~\ref{Figure8}, the trajectory stays a long time close to
its initial position and then make a fast
excursion before an equally fast return to the departure point.
The fast oscillation corresponds to an important exchange of energy between the
carrier and the side bands since the carrier almost reaches its maximal energy
during the burst.
 We see in the Fig.~\ref{Figure9}
the shape of the wave in this case which is quasiperiodic.
\end{itemize}

\section{Dynamics with two resonances}
\label{sec4}
In this section we analyze the dynamics of Hamiltonian~(\ref{504}) with the two
resonances. The phase space is now the product of $\mathbb{E}_{+}$ by $\mathbb{E}_{-}$
associated with each resonance.
 In this case, the dynamics shows a
rich variety of regimes allowing a more efficient energy transfer among the
modes. We give the dynamics of different orbits in the phase space for different values of
the parameters and also give   the type
of the shape of the                                                                                                                                                                                                                                                                                                                                                                                                                                              wave $\eta(x,t)$.  \\
As in the case of one resonance, we start by simple situations: we explore the dynamics
of the points $P_+ \times  P_-$, $S_+ \times  S_-$, $T_+ \times  T_-$ and the segment
$(P_+S_+) \times  (P_-S_-)$ for which we get
the analytical expression of the wave $\eta(x,t)$.\\

\subsection{Dynamics on points $P_+ \times  P_-$, $S_+ \times  S_-$ and $T_+ \times  T_-$}
\label{sec41}
Let us
start with
 the case where the position of the projection of the orbits
on  $(\mathbb{E}_{+}) \times (\mathbb{E}_{-})$ is fixed at $P_+ \times  P_-$
which correspond to $\mathbb{I}_+=\mathbb{I}_-=0$ (see Fig.~\ref{Figure1}).
Since $\mathbb{I}_{+}=2N_1+N_3$ is a conserved quantity and $N_1$, $N_3$ are such that
 $N_1 \geq 0$ and $N_3 \geq 0$, we have
 $N_1=N_3=0$ for all time and $N_2=\mathbb{J}_+/3$.
Similarly we have $N_{-1} =N_{-3} = 0$ and  $N_{-2}=\mathbb{J}_+/3$. \\
Then for $P_+ \times  P_-$, all the actions are constant and the dynamics is connected
to the evolution of the angles. The
corresponding dynamical equations are:
\begin{eqnarray}
i\frac{db_{2}(t)}{dt}&=&(3+2U_2(N_2-2N_{-2}))b_{2}, \\
i \frac{db_{-2}(t)}{dt}&=&(3+2U_2(N_{-2}-2N_2))b_{-2}. \nonumber\\
\label{597}
\end{eqnarray}
The solution in $b_j(t)$ is then :
\begin{eqnarray}
 b_{2}(t)&=& \sqrt{\mathbb{J}_+/3} \exp\left( -i( {9ct + \phi_{+}})\right) , \nonumber\\
 b_{-2}(t)&=&\sqrt{\mathbb{J}_+/3} \exp\left(-i ({9 \alpha ct + \phi_{-}})\right),  \nonumber \\
 N_j(t)&=&0, \quad \mbox{for} \quad j=1,-1,3,-3,
\end{eqnarray}
where
 \begin{equation}
\label{612} \alpha=\frac{J_0+\mathbb{J}_{-}-2\mathbb{J}_{+}}
{J_0+\mathbb{J}_{+}-2\mathbb{J}_{-}},
\quad   c=\frac{1}{3} (1 + \frac{\mathbb{J}_{+} - 2 \mathbb{J}_{-}}{\mathbb{J}_{0}}),  \quad
 \mbox{and}  \quad
\mathbb{J}_{0}=\frac{9}{2U_2}.
\end{equation}
We notice that $\alpha$  is related to $\mathbb{J}_{+}$ and $\mathbb{J}_{-}$ by
\begin{equation}
(1+2\alpha) \mathbb{J}_{-} -
(\alpha+2)\mathbb{J}_{+}=(\alpha-1)\mathbb{J}_0. \nonumber
\end{equation}
The expression of the shape $\eta(x,t)$ is given by:
\begin{eqnarray}
\label{558} \eta (x,t)= \frac{g^{-1/4}}{2\pi \sqrt{2}}\sum_{j=1,2,3} {\vert
k_j\vert} ^{1/4} \biggl( (a_j+a_{-j}^{*}) e
^{ik_jx}+(a_j^{*}+a_{-j}) e^{-ik_jx} \biggr),
\end{eqnarray}
where the  variables $a_j$, as function of  $b_j$ are given by
(see Remark 2 of the Appendix):
\begin{eqnarray}
\label{557}
a_{j}&=&b_{j}+3\widetilde{V}^{(4)}_{{j},{j},{-j},{-j}}b_{-j}^{*2}b_{j}^{*},
\qquad \mbox{for} \quad j=1,2,3.
\end{eqnarray}
Similarly $a_{-j}$ is given by  replacing $k_j$ by $-k_j$ in Eq. (\ref{557}), and $a_{j}=0$
for $j\neq 2$, $j\neq -2$.
Then we deduce the surface shape:
\begin{eqnarray}
\label{559}
\eta(x,t)&=& \frac{g^{-1/4}}{\pi \sqrt{2}}
\biggl[\sqrt{\mathbb{J}_{+}}
\cos(9(x-ct)-\phi_{+})+\sqrt{\mathbb{J}_{-}}\cos(9(x+\alpha ct)+\phi_{-})\nonumber\\ &&
+  V \mathbb{J}_{-}\sqrt{\mathbb{J}_{+}}\cos(9(x+(2\alpha+1) ct)+ 2\phi_{-}+\phi_{+})
\nonumber\\ &&
+V \mathbb{J}_{+}\sqrt{\mathbb{J}_{-}}\cos(9(x-(\alpha+2) ct)- 2\phi_{+}-\phi_{-})\biggr],
\nonumber\\
\end{eqnarray}
where $ V = \widetilde{V}^{(4)}_{{2},{2},{-2},{-2}}\approx -0.082$.\\
The wave is the sum of four traveling waves with four a priori different velocities
$c, \alpha c, (2\alpha+1)c$, and $(\alpha+2)c$.\\
According to the values of $\mathbb{J}_{+}$
 and $\mathbb{J}_{-}$ and more precisely, depending on $\alpha$, these solutions are
fixed points, traveling or standing waves, periodic or quasiperiodic
solutions.\\

\begin{itemize}
\item  If $\mathbb{J}_{+}=\mathbb{J}_{-}$, then $\alpha=1$. We deduce the shape of the wave~:
\begin{eqnarray}
\eta(x,t)&=& \frac{g^{-1/4}\sqrt{2}}{\pi}
\sqrt{\mathbb{J}_{+}}  \cos \biggl(9x-\frac{\phi_{+}-\phi_{-}}{2} \biggr) \nonumber\\ &&
 \times \biggl(\cos(9ct+\frac{\phi_{+}+\phi_{-}}{2})
+ V  \mathbb{J}_{+}  \cos(27ct+3 \frac{\phi_{+}+\phi_{-}}{2})
  \biggr). \nonumber
\end{eqnarray}
which is a standing wave. Such wave is represented in Fig.~\ref{Figure10}(a). \\
\item For the particular case $\mathbb{J}_{+}=\mathbb{J}_{-}=J_0$, $\alpha$ is not defined
and  the velocities  {\it c} and $\alpha c$  vanish. The expression of the surface
wave is:
\begin{eqnarray}
\eta(x,t)&=& \frac{g^{-1/4}\sqrt{2}}{\pi}
\sqrt{\mathbb{J}_{+}}  \biggl(\cos(9x-\frac{\phi_{+}-\phi_{-}}{2} \biggr) \nonumber\\ &&
 \times \biggl(\cos(\frac{\phi_{+}+\phi_{-}}{2})
+ V  \mathbb{J}_{+}  \cos(3 \frac{\phi_{+}+\phi_{-}}{2})
  \biggr). \nonumber
\end{eqnarray}
 The wave is  stationary  (see Fig.~\ref{Figure10}(b)) and its  amplitude is a function
  of the initial phases.\\

\item When $\mathbb{J}_{+}+\mathbb{J}_{-}=2J_0$ so that $\alpha=-1$, we have the unique
family of traveling waves with velocity {\it c} whose shape is given by:
\begin{eqnarray}
\eta(x,t)&=& \frac{g^{-1/4}}{\pi \sqrt{2}}
\biggl[\sqrt{\mathbb{J}_{+}}
\cos(9(x-ct)-\phi_{+})+\sqrt{\mathbb{J}_{-}}\cos(9(x- ct)+\phi_{-})\nonumber\\ &&
+  V \mathbb{J}_{-}\sqrt{\mathbb{J}_{+}}\cos(9(x- ct)+ 2\phi_{-}+\phi_{+})
\nonumber\\ &&
+V \mathbb{J}_{+}\sqrt{\mathbb{J}_{-}}\cos(9(x-ct)- 2\phi_{+}-\phi_{-})  \biggr].
\end{eqnarray}
Fig.~\ref{Figure10}(c) shows the elevation of this surface.\\
\item Time periodic waves  are obtained for $\alpha \in \mathbb{Q}$ including the particular
case $-\mathbb{J}_{+}+2\mathbb{J}_{-} =J_0$ and $-\mathbb{J}_{-}+2\mathbb{J}_{+} =J_0$,
for which
$ \alpha$  is respectively indefinite or zero.

The ratios between the four angles,
 evolving in time,
 which are  $-ct$, $\alpha ct$, $(2\alpha+1)ct$  and $-(\alpha+2)ct$ are
 rational.
The wave is periodic both in space and time, forming the square-shape
patterns as observed in Fig.~\ref{Figure10}(d).\\
\item  In the  case where  $\alpha$ is irrational, the wave is quasiperiodic in time,
expressed as a sum of four traveling waves with different velocities.
We have plotted such wave in Fig.~\ref{Figure10}(e).\\
\end {itemize}
We notice that a similar variety of regular surface shapes
( standing, stationary, traveling, time-periodic and quasiperiodic waves)
are obtained for the two other points of $ (\mathbb{E}_{+}) \times (\mathbb{E}_{-})$
which are $ S_{+}\times S_{-}$, $ T_{+}\times T_{-}$, and for the points $ P_{+}\times S_{-}$,
$ P_{+}\times T_{-}$, $ S_{+}\times T_{-}$, $ S_{+}\times P_{-}$,
$ T_{+}\times S_{-}$, and  $ T_{+}\times P_{-}$.\\

\subsection{Dynamics of  a point on the segment $(P_+S_{+}) \times  (P_-S_{-})$}
\label{sec42}
We consider the point which is at the intersection of
$ C_{+}(\mathbb{I}_{+}) \times C_{-}(\mathbb{I}_{-})$ and $(P_+S_{+}) \times  (P_-S_{-})$.
This point is such that $b_3(0)=0$ and $b_{-3}(0)=0$. From equations of motion of
Hamiltonian (\ref{504}), we deduce $b_3(t)=0$ and $b_{-3}(t)=0$ for any time.
From the invariants we obtain that
 $N_1$, $N_2$, $N_{-1}$ and $N_{-2}$ are constant and equal to $N_1(t)=\mathbb{I}_+/2$
and $N_2(t)=(2\mathbb{J}_+-\mathbb{I}_+)/6$, $N_{-1}(t)=\mathbb{I}_-/2$ and
$N_{-2}(t)=(2\mathbb{J}_--\mathbb{I}_-)/6$ .
The equations of motion become:
\begin{eqnarray*}
\frac{d{b_1}}{dt}&=&-i\left( 1+ (U_1-\frac{V_{21}}{6}) \mathbb{I}_+
-  (2U_1+\frac{V_{-21}}{6}) \mathbb{I}_-
 + \frac{V_{21}}{3}\mathbb{J}_+ +\frac{V_{-21}}{3}\mathbb{J}_-\right) b_1, \\ \nonumber
\frac{d{b_2}}{dt}&=&-i\left( 3+ (-\frac{U_2}{3}+\frac{V_{21}}{2}) \mathbb{I}_+
+ (\frac{2U_2}{3}+\frac{V_{-21}}{2}) \mathbb{I}_-
 + \frac{2}{3} U_2\mathbb{J}_+ - \frac{4}{3}U_2 \mathbb{J}_-\right) b_2,\\ \nonumber
\frac{d{b_{-1}}}{dt}&=&-i\left( 1 -  (2U_1+\frac{V_{-21}}{6}) \mathbb{I}_+
+(U_1-\frac{V_{21}}{6}) \mathbb{I}_-
+\frac{V_{-21}}{3}\mathbb{J}_+ + \frac{V_{21}}{3}\mathbb{J}_- \right) b_{-1}, \\ \nonumber
\frac{d{b_{-2}}}{dt}&=&-i\left( 3 + (\frac{2U_2}{3}+\frac{V_{-21}}{2}) \mathbb{I}_+
+(-\frac{U_2}{3}+\frac{V_{21}}{2}) \mathbb{I}_-
- \frac{4}{3}U_2 \mathbb{J}_+ + \frac{2}{3} U_2\mathbb{J}_- \right) b_{-2}.
\end{eqnarray*}

We deduce the expression of the shape of the wave :
\begin{eqnarray}
\eta(x,t)&=&\frac{g^{-1/4}}{{2} \pi}
 \biggl[ \sqrt{\mathbb{I}_+} \cos(x+c_1t+\alpha_1)
 + \sqrt{{2\mathbb{J}_+ -\mathbb{I}_+}} \cos(9(x-c_2 t)+\alpha_2)  \nonumber \\
&& + \sqrt{\mathbb{I}_-} \cos(x+c_{-1}t+\alpha_{-1})
 + \sqrt{{2\mathbb{J}_- -\mathbb{I}_-}} \cos(9(x-c_{-2} t)+\alpha_{-2}) \biggr]
 , \nonumber \\
 \label{104}
\end{eqnarray}
where  $c_1=1+ (U_1+{V_{21}}/6) \mathbb{I}_+ -  (2U_1+{V_{-21}}/6) \mathbb{I}_-
+ {V_{21}\mathbb{J}_+}/3 +{V_{-21 }\mathbb{J}_-}/{3}$,\\
$c_2=\left[ 3+ (-{U_2}/{3}+{V_{21}}/{2}) \mathbb{I}_+
+ ({2U_2}/{3}+{V_{-21} }/{2})\mathbb{I}_-
+ {2} U_2 \mathbb{J}_+/{3} - {4}U_2 \mathbb{J}_-/{3}\right]/9$,\\
 $c_{-1}=1-  (2U_1+{V_{-21}}/{6})\mathbb{I}_+  + (U_1+{V_{21}}/6) \mathbb{I}_-
+{V_{-21}\mathbb{J}_+}/{3} + {V_{21}\mathbb{J}_-}/3 $, and  \\
$c_{-2}=\left[ 3 + ({2U_2}/{3}+{V_{-21}}/{2}) \mathbb{I}_+
+ (-{U_2}/{3}+{V_{21}}/{2})\mathbb{I}_-
- {4}U_2 \mathbb{J}_+/ {3}+ {2} U_2\mathbb{J}_-/{3} \right]/9$.\\
This wave is the sum of four  traveling waves with four different velocities. \\

\subsection{Dynamics of  orbits on the segment $(P_+S_{+})$ times the interior of
$C_{-}(\mathbb{I}_{-})$}
\label{sec43}
We consider the dynamics of $(P_+S_{+})$ times  the interior of $C_{-}(\mathbb{I}_{-})$
(or similarly the interior of $C_{+}(\mathbb{I}_{+})$ times $(P_-S_{-})$).
We have $b_3(t)=0$ for all $t$, so
$N_3(t)=0$ is a constant of motion and $N_1(t)$ and  $N_2(t)$ are constant for all t.
We have now a system with five degrees of freedom  $b_{1}$, $b_{2}$, $b_{-1}$,
$b_{-2}$ and $b_{-3}$
with five constants of motion,
$\mathbb{I}_{+}$, $\mathbb{I}_{-}$, $\mathbb{J}_{+}$, $\mathbb{J}_{-}$ and $H$.
 It is then integrable. The modes $b_1$ and $b_2$ evolve with a constant modulus and
 a phase depending on $\vert b_{-1}\vert$, $\vert b_{-2}\vert$ and $\vert b_{-3}\vert$.
The dynamics of the modes $b_{-1}$, $b_{-2}$ and $b_{-3}$ in $C_{-}(\mathbb{I}_{-})$
depend on the second resonance just through the modulus of  $b_{1}$ and  $b_{2}$ which are constant.
So the exchanges of energy on $C_{-}(\mathbb{I}_{-})$ are not influenced by the evolution of
$b_{1}$ and  $b_{2}$. The dynamics  of the modes $b_{-1}$, $b_{-2}$ and $b_{-3}$ is then similar
to the one studied in the case with one resonance in Sec.~\ref{sec3}.\\

\subsection{Dynamics in the interior of
$ C_{+}(\mathbb{I}_{+}) \times C_{-}(\mathbb{I}_{-})$}
\label{sec44}
In order to study the dynamics on the nontrivial sets
$ C_{+}(\mathbb{I}_{+}) \times C_{-}(\mathbb{I}_{-})$ on the interior of the phase space
$ (\mathbb{E}_{+}) \times (\mathbb{E}_{-})$, i.e. such that $N_j$ are strictly positive for
$j=-1,-2,-3,1,2,3$, we perform a canonical transformation in action-angle coordinates,
$b_j=\sqrt{N_j} e^{-i\Phi_j}$ and  Hamiltonian (\ref{504}) becomes:

\begin{eqnarray}
H&=&\sum_{j=1,2,3}
 \omega_{j}(N_j+N_{-j}) \nonumber
 +\sum_{j=1,2,3}U_{j}(N_j^2+ N_{-j}^2-4N_j N_{-j}) \nonumber \\
 && +
  V_{21}(N_1 N_2+N_{-1}N_{-2}) + V_{-21}(N_1 N_{-2}+N_{-1}N_2) \nonumber \\
 && - V_{31}(N_1N_3+N_{-1}N_{-3}) + V_{-31}(N_1N_{-3}+N_{-1}N_3) \nonumber \\
 && +
 V_{23}(N_2N_3+N_{-2}N_{-3}) - V_{-23}(N_{2}N_{-3}+N_{-2}N_3)\nonumber \\
 &&+W\sqrt{N_1N_2}N_3\cos (\Phi_1+\Phi_2-2\Phi_3) \nonumber \\
 && +W\sqrt{N_{-1}N_{-2}}N_{-3}\cos (\Phi_{-1}+\Phi_{-2}-2\Phi_{-3}) . \label{HamAA}
\end{eqnarray}

Using the invariants $\mathbb{I}_+$, $\mathbb{I}_-$, $\mathbb{J}_+$ and $\mathbb{J}_-$
, we now proceed
to the reduction of the dimension of
the system to two degrees of freedom.\\
We perform as in the case with one resonance
the  canonical transformation from \\
$(N_1,N_2,N_3,N_{-1},N_{-2},N_{-3},\Phi_1,
\Phi_2,\Phi_3,\Phi_{-1},\Phi_{-2},\Phi_{-3})$ to \\
 $(\tilde{N_1},\mathbb{K}_+,\mathbb{I}_+, \tilde{N_{-1}},\mathbb{K}_-,\mathbb{I}_-,
\theta_1,\theta_2,\theta_3,\theta_{-1},\theta_{-2},\theta_{-3})$:

\begin{eqnarray}
\tilde{N_1}&=&N_1, \nonumber  \\
\mathbb{K}_+&=&N_1-N_2, \nonumber  \\
\mathbb{I}_+&=&2N_1+N_3, \nonumber  \\
\theta_1&=&\Phi_1+\Phi_2-2\Phi_3, \nonumber  \\
\theta_2&=&-\Phi_2, \nonumber    \\
\theta_3&=&\Phi_3, \nonumber    \\
\tilde{N}_{-1}&=&N_{-1}, \nonumber  \\
\mathbb{K}_-&=&N_{-1}-N_{-2}, \nonumber  \\
\mathbb{I}_-&=&2N_{-1}+N_{-3}, \nonumber  \\
\theta_{-1}&=&\Phi_{-1}+\Phi_{-2}-2\Phi_{-3}, \nonumber  \\
\theta_{-2}&=&-\Phi_{-2}, \nonumber    \\
\theta_{-3}&=&\Phi_{-3}.
\end{eqnarray}
 The actions $\tilde{N_1}$,
$\mathbb{K}_+$, $\mathbb{I}_+$, $\tilde{N}_{-1}$, $\mathbb{K}_-$ and $\mathbb{I}_-$
are now respectively conjugate to angles
$\theta_1$, $\theta_2$,  $\theta_3$, $\theta_{-1}$,
$\theta_{-2}$, and $\theta_{-3}$. In what follows, we use $N_1$, $N_{-1}$
instead of $\tilde{N}_1$ and $\tilde{N}_{-1}$
for simplicity. Hamiltonian~(\ref{HamAA}) becomes:

\begin{eqnarray}
H
&=&\Omega_{0}
+\Omega_{1}N_{1}+\Omega_{-1}N_{-1}
+\Omega_{2}(N_{1}^{2}+N_{-1}^{2})+\Omega_{3}N_{1}N_{-1} \nonumber \\
&&+W\sqrt{N_{1}}\sqrt{N_{1}-\mathbb{K}_{+}}(\mathbb{I}_{+}-2N_{1})\cos
\theta _{1} \nonumber \\
&&+W\sqrt{N_{-1}}\sqrt{N_{-1}-\mathbb{K}_{-}}(\mathbb{I}_{-}-2N_{-1})\cos
\theta _{-1}, \label{541}
\end{eqnarray}
where the functions $\Omega_0$, $\Omega_1$ and $\Omega_{-1}$ only depend
on the actions $\mathbb{I}_{+}$, $\mathbb{K}_{+}$, $\mathbb{I}_{-}$,
$\mathbb{K}_{-}$ and $\Omega_2$ and $\Omega_3$ are constant~:
\begin{eqnarray}
\Omega_{0}&=&-3(\mathbb{K}_{-}+\mathbb{K}_{+})+2(\mathbb{
I}_{-}+\mathbb{I}_{+})
+U_{3}(\mathbb{I}_{-}^{2}+\mathbb{I}_{+}^{2}-4\mathbb{I}_{+}\mathbb{I}_{-})
\nonumber \\
&&+U_{2}(\mathbb{K}_{-}^{2}+\mathbb{K}
_{+}^{2}-4\mathbb{K}_{+}\mathbb{K}_{-}) \nonumber \\
&&-V_{23}(\mathbb{I}_{+}\mathbb{K}_{+}+\mathbb{I}_{-}\mathbb{K}_{-})+
V_{-23}(\mathbb{I}_{+}\mathbb{K}
_{-}+\mathbb{I}_{-}\mathbb{K}_{+}), \nonumber\\
\Omega_{1}
&=&2U_{2}(-\mathbb{K}
_{+}+2\mathbb{K}_{-})+4U_{3}(-\mathbb{I}_{+}+2\mathbb{I}_{-})-V_{21}\mathbb{K}_{+}
-V_{-21}\mathbb{K}_{-} \nonumber \\
&&-V_{31}\mathbb{I}_{+} + V_{-31} \mathbb{I}_{-} +
V_{23}(\mathbb{I}_{+}+2\mathbb{K}_{+})
-V_{-23}(\mathbb{I}_{-}+2\mathbb{K}_{-}), \nonumber \\
\Omega_{-1}
&=&2U_{2}(-\mathbb{K}
_{-}+2\mathbb{K}_{+})+4U_{3}(-\mathbb{I}_{-}+2\mathbb{I}_{+})-V_{21}
\mathbb{K}_{-}-V_{-21}\mathbb{K}_{+} \nonumber \\
&&-V_{31} \mathbb{I}_{-}+
V_{-31}\mathbb{I}_{+}+V_{23}(\mathbb{I}_{-}+2\mathbb{K}_{-})-V_{-23}(
\mathbb{I}_{+}+2\mathbb{K}_{+}), \nonumber \\
\Omega_{2}
&=&U_{1}+U_{2}+4U_{3}+V_{21}+2V_{31}-2V_{23}, \nonumber \\
\Omega_{3}
&=&-4U_{1}-4U_{2}-16U_{3}+2V_{-21}-4V_{-31}+4V_{-23}.
\end{eqnarray}
Numerically  $\Omega_{2}$  and $\Omega_{3}$ are given by:
$\Omega_{2}=-3.9642$,
$\Omega_{3}=-16.0933$. \\
Since Hamiltonian~(\ref{541}) does not depend on $\theta_2$,
$\theta_3$, $\theta_{-2}$ and $\theta_{-3}$, we can consider
$\mathbb{I}_{+}$, $\mathbb{K}_{+}$,
$\mathbb{I}_{-}$, $\mathbb{K}_{-}$ as parameters of the system for
the dynamics of $N_1$, $N_{-1}$, $\theta_{1}$ and $\theta_{-1}$.\\
We recall that $N_1$ and  $N_{-1}$ are such that
$\rm{max}(0,\mathbb{K}_{+}) < N_1 < \mathbb{I}_{+}/2$ and
$\rm{max}(0,\mathbb{K}_{-}) < N_{-1} < \mathbb{I}_{-}/2$.
The accessible part of phase space is thus contained into the product of
two interiors of a circle or an annulus, depending on the values of $\mathbb{I}_{+}$
and $\mathbb{I}_{-}$.\\
In order to visualize the dynamics on $ C_{+}(\mathbb{I}_{+}) \times C_{-}(\mathbb{I}_{-})$
we compute Poincar\'e sections of the dynamics and then we project these
surfaces on
$(N_{1},\theta_{1})$-plane. The Poincar\'e sections $\Sigma$ are constructed
 in the following way:
 \begin{equation}
 \label{701}
  \Sigma =\left\lbrace (N_{1},N_{-1},\theta_{1},\theta_{-1}): \theta_{-1}=0,
    \frac{d\theta_{-1}}{dt} <0, H=H(N_{1},N_{-1},\theta_{1},\theta_{-1},\mathbb{I}_{+},
    \mathbb{I}_{-})
  \right\rbrace.
\end{equation}
Now the parameters are the energy $H$, and the four invariants
$ \mathbb{I}_{+} $,
$ \mathbb{I}_{-} $, $ \mathbb{J}_{+} $ and $ \mathbb{J}_{-} $. \\
Since Hamiltonian (\ref{HamAA}) has two degrees of freedom, Poincar\'e sections
are two dimensional and therefore we get an accurate picture of the dynamics by looking at the
projections of $\Sigma$ on $(N_{1},\theta_{1})$. For most of the values of the parameters,
these sections show a rich variety for the dynamics, a mixed phase space with regular and
chaotic behaviors.\\
We give a plot of some projections  for  $ \mathbb{J}_{+} =\mathbb{J}_{-}=1$,
$H=2$  and for different values of
$ \mathbb{I}_{+}=\mathbb{I}_{-}$: from $ \mathbb{I}_{+}=\mathbb{I}_{-}=0.3$ to
$ \mathbb{I}_{+}=\mathbb{I}_{-}=0.8$.\\
We observe on Fig.~\ref{Figure11}(a), when  $ \mathbb{I}_{+}$ and $ \mathbb{I}_{-}$
are near zero so that
 we are in the
neighborhood of the  point $P_+ \times P_-$, the motion is completely regular and all
the orbits lie on invariant tori. For this value of $H$, the dynamics occurs close to
the center of the accessible region of phase space.
The elliptic and  hyperbolic fixed points of
the Poincar\'e sections are of the same type as those of the elliptic and
hyperbolic orbits of the case with one resonance.
Besides these points,
there are one elliptic and two hyperbolic points due to the second resonance.
We recall that the invariants
depend on the variables of only one resonance. So, for small values of $\mathbb{I}_{+}$
and $\mathbb{I}_{-}$
we can make the same reasoning as for the one resonance case and we can conclude that
in the neighborhood of the
point  $P_+ \times P_-$, the exchanges of energy between the modes are
not significant and it leads to a regular behavior of the system.
Thus in the system with small values of  $ \mathbb{I}_{+}$ and $\mathbb{I}_{-}$,
the only possible motion for the system
is periodic or quasi-periodic in time, corresponding to regular shapes for the wave $\eta(x,t)$
(see Sec.~\ref{sec3}).
\\ To undergo a transition to chaos,
we increase the parameter
 $\mathbb{I}_{+}$ ( which increases the effect of the nonlinear terms up to
a certain threshold),
and for a value of $\mathbb{I}_{+}$ in the neighborhood of $\mathbb{I}_{+}=\mathbb{J}_{+}/2$,
we see in Fig.~\ref{Figure11}(b) and Fig.~\ref{Figure11}(c) , that the phase space is made up of regular and
chaotic motions. Chaotic trajectories are located between
regular trajectories (on invariant tori). We observe
that the motion is clearly affected by the vicinity of
hyperbolic points: the irregular motion appears in the neighborhood of  hyperbolic
periodic orbits breaking up their neighboring KAM tori.\\
The first of the two regular zones which surround the chaotic sea (outer region),
  is obtained  for large amplitudes
of the modes of low frequency $N_{1}$ and
$N_{-1}$ which are larger than the ones  of  high frequency modes
$N_{3}$ and $N_{-3}$.
The second regular zone (inner region) is obtained when the amplitude
of the modes of low frequency $N_{1}$ and
$N_{-1}$ are small. In these
regular zones, the exchanges of energy are not significant and give rise to
regular behaviors of the system. On the other hand, for orbits close to the hyperbolic points,
the modes oscillate chaotically (see Fig.~\ref{Figure12}). As the parameter
$\mathbb{I}_{+}$  increases, the region in which
$N_{1}$ evolves shrinks and all the orbits become stable, the system
admits only regular behaviors as it can be seen on  Fig.~\ref{Figure11}(d). This is due
to the fact that in the
accessible region, the values  of $N_{1}$ are larger than those of $N_{3}$ and
for this case the exchanges among the modes are not significant.\\
We conclude that chaotic exchanges of energy may be done in the neighborhood of
$ C_{+}(\frac{\mathbb{J}_{+}}{2})
\times C_{-}(\frac{\mathbb{J}_{-}} {2})$.
In Fig.~\ref{Figure12}, we give the plot of the modes $N_j(t), j=1,2,3,-1,-2,-3$ and in
 Fig.~\ref{Figure13}, we give
the shape of the wave $\eta(x,t)$
for $\mathbb{I}_{+}=\mathbb{I}_{-}=\mathbb{J}_{+}/2$ and
for an initial condition $(N_1,\theta_1)=(0.152,-3.07)$ which is in the chaotic region.
In these two plots, we see the irregular motion of the modes $N_j$ (chaotic exchanges of energy)
, and the chaotic
structure of  $\eta(x,t)$.
We notice that in the chaotic zones, significant energy exchanges occur among the modes.

\section{Conclusion}
Following the framework of Zakharov in  Ref.~\cite{Za} and  Krazitskii  in Ref.~\cite{Kra},
we have built a Hamiltonian model for six interacting waves for gravity water waves
in infinite depth. The interaction are those of two coupled Benjamin-Feir resonances.
The knowledge of four constants of motion allowed us to reduce  the
system to a Hamiltonian system with two degrees of freedom. We have first
studied the dynamical properties
of the Hamiltonian with only one resonance which is  integrable. This simple
case allowed us to understand the basics of the full dynamics. A whole variety
of solutions have been found in this simple case such as traveling, time periodic and quasi-periodic waves.
 We showed that exchanges of energy occur in the zone of  phase space where
 the energies initially put into the two side bands

  are equally important and when the energy  in the central
 mode is the same as the sum of the energies of the two side bands. This situation corresponds
 to phase space regions close to the separatrices.
  All these exchanges are done in a periodic way.
  Thus, when we take into account only one resonance, the only possible motion
  is a regular time quasiperiodic (and space periodic) wave.\\
  Then we have studied the dynamics with the two resonances. Besides the regular waves such as
  traveling, time periodic and quasiperiodic waves,
  the full system exhibits
  stationary and standing waves.
  We have visualized the dynamics using  numerical Poincar\'e sections.
  We have showed that the system with the two coupled resonances undergoes a
  transition to chaos. Chaotic behaviors occur  in the regions
  where the energies of the side bands of each resonance are equally important and near
  the broken separatrices.
  In this case, there is a strong exchange of energy between all the modes and it is done in chaotic way.

\appendix

\section*{Derivation of the Hamiltonian given by Eq. (\ref{504})}

In this appendix, we briefly derive the expression of Hamiltonian (\ref{504}) from the Hamiltonian
derived
 by Krazitskii  in
Ref.~\cite{Kra} given by Eq. (2.11). We get the expression for the Hamiltonian in the case
where the three modes (denoted by 1, 2 and 3)
 are such that $ k_1+k_2=2k_3$ and $\omega_1+\omega_2=2\omega_3$.\\ In this case, we notice
 that there are no terms of order 3, i.e.
terms of the form $ a_k a_{k^{'}} a_{k^{''}} $  such that $ k= k^{'}+k^{''}$
for $ k,k^{'},k^{''}$ $\in$
$\{k_1,k_2,k_3,-k_1,-k_2,-k_3 \}$. \\
We expand this Hamiltonian   up to order four,  i.e. we neglect terms of order
five in Hamiltonian
(2.11) of  Ref.~\cite{Kra} since we are in the case of weak nonlinearity
( small amplitudes of the wave).
 Hamiltonian (2.11) of Ref.~\cite{Kra} becomes:
\begin{eqnarray}
\label{E1}
H &=&\sum_{i=1,2,3}\omega
_{i}(a_{i}a_{i}^{*}+a_{-i}a_{-i}^{* }) +
\nonumber \\ &&
 +6\sum_{i,j=1,2,3,\atop i\neq j } V_{i,i,-j,j}^{(1)}(a_{-i}^{
* }a_{-i}a_{-j}a_{j}+a_{i}^{* }a_{i}a_{j}a_{-j}+c.c) \nonumber
\\
&& + 3V_{2,-1,3,3}^{(1)}(a_{2}^{*}a_{-1}a_{3}^{2}+a_{-2}^{*
}a_{1}a_{-3}^{2}+c.c)
+3V_{1,-2,3,3}^{(1)}(a_{1}^{*}a_{-2}a_{3}^{2}+a_{-1}^{*
}a_{2}a_{-3}^{2}+c.c)
\nonumber \\ &&\
+3V_{3,-3,1,2}^{(1)}(a_{3}^{*}a_{-3}a_{1}a_{2}+a_{-3}^{*
}a_{3}a_{-1}a_{-2}+c.c) \nonumber \\
&&+\frac{1}{2}\sum_{i=1,2,3}V_{i,i,i,i}^{(2)}(a_{i}^{*
2}a_{i}^{2}+a_{-i}^{\ast 2}a_{-i}^{2})
+2\sum_{i=1,2,3}V_{i,-i,i,-i}^{(2)}(a_{i}^{*
}a_{-i}^{*}a_{i}a_{-i}) \nonumber \\ &&\
+2\sum_{i,j=1,2,3,\atop i\neq j}
\biggl(V_{i,j,i,j}^{(2)}(a_{i}^{*}a_{j}^{*}a_{i}a_{j}+a_{-i}^{*
}a_{-j}^{\ast }a_{-i}a_{-j})
+V_{i,-j,i,-j}^{(2)}(a_{i}^{* }a_{-j}^{*
}a_{i}a_{-j}+a_{-i}^{\ast }a_{j}^{\ast }a_{-i}a_{j})
\biggr)
\nonumber \\ &&\
+2\sum_{i,j=1,2,3,\atop i\neq j}
V_{-i,i,-j,j}^{(2)} (a_{-i}^{ \ast }a_{i}^{\ast }a_{-j}a_{j} +c.c)
+V_{1,2,3,3}^{(2)}(a_{1}^{\ast }a_{2}^{*
}a_{3}^{2}+a_{-1}^{\ast }a_{-2}^{\ast }a_{-3}^{2}+c.c)
\nonumber \\ &&\
+2V_{3,-1,-3,2}^{(2)}(a_{3}^{\ast
}a_{-1}^{* }a_{-3}a_{2}+a_{-3}^{\ast }a_{1}^{\ast
}a_{3}a_{-2}+c.c) \nonumber \\ &&\
+ \frac{3}{2} \sum_{i=1,2,3}V_{-i,-i,i,i}^{(4)}(a_{-i}^{\ast 2}a_{i}^{*2}+c.c)
+6\sum_{i,j=1,2,3, \atop i\neq j}
V_{-i,i,-j,j}^{(4)}(a_{-i}^{
* }a_{i}^{\ast }a_{-j}^{\ast }a_{j}^{* }+c.c)
\nonumber \\
&&+3V_{1,2,-3,-3}^{(4)}(a_{-1}^{*}a_{-2}^{* }a_{3}^{*
2}+a_{1}^{\ast }a_{2}^{* }a_{-3}^{* 2}+c.c) ,
\end{eqnarray}
where the coefficients verify the symmetry  $V_{i,j,k,l}^{(n)}=V_{-i,-j,-k,-l}^{(n)}$,
for all $n=1,2,4$ and
$i,j,k,l \in \{1,2,3,-1,-2,-3 \}$ as well as  the other symmetries mentioned
in Ref.~\cite{Kra}.
 We  verify in the expressions of
the coefficients
given in  Ref.~\cite{Kra} that  all  the coefficients $V_{i,i,-i,i}^{(1)}$ are equal to zero.

The canonical transformation from the variables $a_{k}$ to the
new canonical variables $b_{k}$, that maps Hamiltonian (\ref{E1})
to its normal form up to terms of order 5
in $b_{j}$ is given by:
 \begin{eqnarray}
 a_{1}&=&b_{1}
\label{E2}
+6\widetilde{V}^{(1)}_{{1},{1},{-2},{2}}(b_{1}b_{2}b_{-2}
+b_{1}b_{2}^{*}b_{-2}^{*})
+6\widetilde{V}^{(1)}_{{1},{1},{-3},{3}}(b_{1}b_{3}b_{-3}
+b_{1}b_{3}^{*}b_{-3}^{*})\nonumber\\&&
+6\widetilde{V}^{(1)}_{{2},{2},{-1},{1}}(b_{-2}b_{-2}^{*}b_{-1}^{*}
+b_{2}b_{2}^{*}b_{-1}^{*})
+6\widetilde{V}^{(1)}_{{3},{3},{-1},{1}}(b_{-3}b_{-3}^{*}b_{-1}^{*}
+b_{3}b_{3}^{*}b_{-1}^{*})\nonumber\\&&
+3\widetilde{V}^{(1)}_{{2},{-1},{3},{3}}b_{-2}b_{-3}^{*2}
+3\widetilde{V}^{(1)}_{{1},{-2},{3},{3}}b_{-2}b_{3}^2
+3\widetilde{V}^{(1)}_{{3},{-3},{1},{2}}b_{3}b_{-3}^{*}b_{2}^{*}
\nonumber\\&&
+2\widetilde{V}^{(2)}_{{-1},{1},{-2},{2}}b_{-1}^{*}b_{-2}b_{2}
+2\widetilde{V}^{(2)}_{{-1},{1},{-3},{3}}b_{-1}^{*}b_{-3}b_{3}
+2\widetilde{V}^{(2)}_{{-1},{3},{-3},{2}}b_{3}b_{-3}^{*}b_{-2}
\nonumber\\&&
+3\widetilde{V}^{(4)}_{{1},{1},{-1},{-1}}b_{-1}^{*2}b_{1}^{*}
+\widetilde{V}^{(4)}_{{-1},{1},{-2},{2}}b_{-1}^{*}b_{-2}^{*}b_{2}^{*}
+\widetilde{V}^{(4)}_{{-1},{1},{-3},{3}}b_{-1}^{*}b_{-3}^{*}b_{3}^{*}
\nonumber\\&&
+3\widetilde{V}^{(4)}_{{-1},{-2},{3},{3}}b_{2}^{*}b_{-3}^{*2},
\end{eqnarray}

\begin{eqnarray}
\label{E3} a_{2}&=&b_{2}
+6\widetilde{V}^{(1)}_{{1},{1},{-2},{2}}(b_{-1}b_{-1}^{*}b_{-2}^{*}
+b_{1}b_{1}^{*}b_{-2}^{*})
+6\widetilde{V}^{(1)}_{{2},{2},{-3},{3}}(b_{2}b_{3}b_{-3}
+b_{2}b_{3}^{*}b_{-3}^{*})\nonumber\\&&
+6\widetilde{V}^{(1)}_{{2},{2},{-1},{1}}(b_{2}b_{1}b_{-1}
+b_{2}b_{1}^{*}b_{-1}^{*})
+6\widetilde{V}^{(1)}_{{3},{3},{-2},{2}}(b_{-3}b_{-3}^{*}b_{-2}^{*}
+b_{3}b_{3}^{*}b_{-2}^{*})\nonumber\\&&
+3\widetilde{V}^{(1)}_{{2},{-1},{3},{3}}b_{-1}b_{3}^2
+3\widetilde{V}^{(1)}_{{1},{-2},{3},{3}}b_{-1}b_{-3}^{*2}
+3\widetilde{V}^{(1)}_{{3},{-3},{1},{2}}b_{1}^{*}b_{3}b_{-3}^{*}
\nonumber\\&&
+2\widetilde{V}^{(2)}_{{-2},{2},{-1},{1}}b_{-2}^{*}b_{-1}b_{1}
+2\widetilde{V}^{(2)}_{{-2},{2},{-3},{3}}b_{-2}^{*}b_{-3}b_{3}
+2\widetilde{V}^{(2)}_{{3},{-1},{-3},{2}}b_{3}b_{-3}^{*}b_{-1}
\nonumber\\&&
+3\widetilde{V}^{(4)}_{{2},{2},{-2},{-2}}b_{-2}^{*2}b_{2}^{*}
+\widetilde{V}^{(4)}_{{-1},{1},{-2},{2}}b_{-1}^{*}b_{1}^{*}b_{-2}^{*}
+\widetilde{V}^{(4)}_{{-2},{2},{-3},{3}}b_{-2}^{*}b_{-3}^{*}b_{3}^{*}
\nonumber\\&&
+3\widetilde{V}^{(4)}_{{1},{2},{-3},{-3}}b_{1}^{*}b_{-3}^{*2},
\end{eqnarray}
\begin{eqnarray}
\label{E4} a_{3}&=&b_{3}
+6\widetilde{V}^{(1)}_{{1},{1},{-3},{3}}(b_{-1}b_{-1}^{*}b_{-3}^{*}
+b_{1}b_{1}^{*}b_{-3}^{*})
+6\widetilde{V}^{(1)}_{{2},{2},{-3},{3}}(b_{-2}b_{-2}^{*}b_{-3}^{*}
+b_{2}b_{2}^{*}b_{-3}^{*})\nonumber\\&&
+6\widetilde{V}^{(1)}_{{3},{3},{-1},{1}}(b_{3}b_{1}b_{-1}
+b_{3}b_{1}^{*}b_{-1}^{*})
+6\widetilde{V}^{(1)}_{{3},{3},{-2},{2}}(b_{3}b_{2}b_{-2}
+b_{3}b_{2}^{*}b_{-2}^{*})\nonumber\\&&
+6\widetilde{V}^{(1)}_{{2},{-1},{3},{3}}b_{2}b_{-1}^{*}b_{3}^{*}
+6\widetilde{V}^{(1)}_{{1},{-2},{3},{3}}b_{1}b_{-2}^{*}b_{3}^{*}
+3\widetilde{V}^{(1)}_{{3},{-3},{1},{2}}(b_{-3}b_{2}b_{1}+b_{-3}b_{-1}^{*}b_{-2}^{*})
\nonumber\\&&
+2\widetilde{V}^{(2)}_{{-3},{3},{-1},{1}}b_{-3}^{*}b_{-1}b_{1}
+2\widetilde{V}^{(2)}_{{-3},{3},{-2},{2}}b_{-3}^{*}b_{-2}b_{2}
+2\widetilde{V}^{(2)}_{{3},{-1},{-3},{2}}(b_{-3}b_{-2}^{*}b_{1}+b_{-3}b_{2}b_{-1}^{*})
\nonumber\\&&
+3\widetilde{V}^{(4)}_{{3},{3},{-3},{-3}}b_{-3}^{*2}b_{3}^{*}
+\widetilde{V}^{(4)}_{{-1},{1},{-3},{3}}b_{-1}^{*}b_{1}^{*}b_{-3}^{*}
+\widetilde{V}^{(4)}_{{-2},{2},{-3},{3}}b_{-2}^{*}b_{2}^{*}b_{-3}^{*}
\nonumber\\&&
+6\widetilde{V}^{(4)}_{{-1},{-2},{3},{3}}b_{-1}^{*}b_{-2}^{*}b_{3}^{*}.
\end{eqnarray}
Similarly $a_{-j}$ is given by  replacing $k_j$ by $-k_j$ in Eqs.(\ref{E2}),
(\ref{E3}),and (\ref{E4}).\\

The coefficients $\widetilde{V}^{(1)}_{{i},{j},{l},{m}}$,
$\widetilde{V}^{(2)}_{{i},{j},{l},{m}}$   and
$\widetilde{V}^{(4)}_{{i},{j},{l},{m}}$ are given by :
\begin{eqnarray}
\widetilde{V}^{(1)}_{{i},{j},{l},{m}}&=&\frac{{V}^{(1)}_{{i},{j},{l},{m}}}{\omega
_{i}-\omega_{j}-\omega _{l}-\omega_{m}},\nonumber\\
\widetilde{V}^{(2)}_{{i},{j},{l},{m}}&=&\frac{{V}^{(2)}_{{i},{j},{l},{m}}}{\omega
_{i}+\omega_{j}-\omega _{l}-\omega_{m}} \quad \mbox{for} \quad \omega
_{i}+\omega_{j} \neq \omega _{l}+\omega_{m},\nonumber\\
\widetilde{V}^{(4)}_{{i},{j},{l},{m}}&=&\frac{{V}^{(4)}_{{i},{j},{l},{m}}}{\omega
_{i}+\omega_{j}+\omega _{l}+\omega_{m}}.
\end{eqnarray}

{\em Remark~1~:}
We notice that when the modes of one resonance are equal to zero,
e.g, $ b_{-j}=0$ for $j=1,2,3$
then
we have $a_j=b_j$  for the second resonance ($j=1,2,3$).\\

{\em Remark~2~:} When two modes of each  resonance are equal to zero e.g $ b_j=b_{-j}=0$ for $j=1,3$
 then
 $a_2=b_2+2\widetilde{V}^{(4)}_{{2},{2},{-2},{-2}}b_{-2}^{*2}b_{2}^{*}$,
where $\widetilde{V}^{(4)}_{{2},{2},{-2},{-2}}=-0.08188$.\\
 The transformation  which maps $\{a _{j}\}$ into $ \{b_{j}\}$
 is canonical i.e. it satisfies
\begin{equation}
\label{E5}
\sum_{j}  i db_{j} \wedge db^{*}_{j} = \sum_{j}  i da_{j} \wedge da^{*}_{j}.
\end{equation}

This transformation maps Hamiltonian (\ref{E1}) into  Hamiltonian (\ref{504})
with the following
values of the coefficients  $U_i$, W and $V_{ij}$.
\begin{center}
\begin{tabular}{|c|c|c|c|}
\hline $U_{1}$ & $U_{2}$ & $U_{3}$ & W \\ \hline 0.0063 &
4.61645 & 0.40528 & 0.52648
\\ \hline
\end{tabular}
\end{center}
\begin{center}
\begin{tabular}{|c|c|c|c|c|c|}
\hline $V_{21}$ & $V_{-21}$ & $V_{31}$ & $V_{-31}$ & $V_{23}$ &
$V_{-23}$ \\ \hline 0.07599 & 0.53193 & 0.02533 & 0.17312 &
5.16738 & 2.12774
\\ \hline
\end{tabular}
\end{center}


\newpage


\begin{figure}
\caption {\label{Figure1}
Representation of the phase space $\mathbb{E}_+$ and the three families
of fixed points found on $(N_1,\theta_1)$ plane.
The location of the elliptic points $e(\mathbb{I}_+)$, $e_1(\mathbb{I}_+$)
is represented by a dashed line.
The location of the hyperbolic points $h_1(\mathbb{I}_+)$
is represented by a dotted line.
The  bifurcation creating the pair of elliptic/hyperbolic
$e_1(\mathbb{I}_+)$/$h_1(\mathbb{I}_+)$ point is indicated by a dot.}

\caption{\label{Figure2}
Elevation of the surface $\eta(x,t)$ corresponding to
$P$ $(\mathbb{I}_{+}=0)$
for Hamiltonian (\ref{100}) with one resonance.}

\caption{\label{Figure3} Phase space trajectories of  Hamiltonian (\ref{100})
 with one resonance for different values of
$ \mathbb{I}_{+}$:
  $ (a)$ $\mathbb{I}_{+}=0.1$,    $ (b)$ $ \mathbb{I}_{+}=0.12$,
   $ (c)$ $ \mathbb{I}_{+}=0.5$,
  $ (d) $ $\mathbb{I}_{+}=0.505$,  $ (e)$ $ \mathbb{I}_{+}=0.6$,
   $(f)$ $ \mathbb{I}_{+}=1.7.$,
The grey region is the accessible region of phase space:
which is the annulus
$\max (0,\mathbb{K}_{+}) \leq N_1 \leq \mathbb{I}_{+}/2.$
The bold curve $\mathcal{T}_1$ represents the trajectory which passes by $N_1=0$}

\caption{\label{Figure4} Elevation of the surface $\eta(x,t)$ for Hamiltonian (\ref{100})
with one resonance for
$\mathbb{J}_{+}=1$, $ \mathbb{I}_{+}={\mathbb{J}_{+}}/{2}$,
and  for the fixed point $e(\mathbb{J}_{+}/2).$}

\caption{\label{Figure5} Temporal evolution of modes $N_{i}$ for $i=1,2,3$,
for Hamiltonian (\ref{102}) with one resonance,
$ \mathbb{I}_{+} =1.2$,
$\mathbb{J}_+=1$ and an initial  condition
$(N_1,\theta_1)=({\mathbb{I}_{+}}/{2},\pi)$.}

\caption{\label{Figure6} Temporal evolution of
modes $N_{i}$ for $i=1,2,3$, for Hamiltonian (\ref{102}) with one resonance
,  $\mathbb{J}_+=1$, $ \mathbb{I}_{+} $ in the neighborhood of
$ \mathbb{I}_{+} =\mathbb{J}_+/2$,
and an initial condition
$(N_1,\theta_1)=(\mathbb{J}_+/8,0)$.}

\caption{\label{Figure7}
Elevation of the surface $\eta(x,t)$  Hamiltonian (\ref{102})
with one resonance for
$\mathbb{J}_{+}=1$, in the neighborhood of $ \mathbb{I}_{+}={\mathbb{J}_{+}}/{2}$,
and and an initial condition  $(N_1,\theta_1)=(\mathbb{J}_+/8,0)$ }

\caption{\label{Figure8} Temporal evolution of
modes $N_{i}$ for $i=1,2,3$, for Hamiltonian (\ref{102}) with one resonance
, $\mathbb{J}_+=1$, $ \mathbb{I}_{+} =\mathbb{J}_+/2 $,
 and an initial condition on the separatrix
$(N_1,\theta_1)=(\mathbb{J}_+/4,2.8)$.}

\caption{\label{Figure9}
Elevation of the surface $\eta(x,t)$ for Hamiltonian (\ref{102})
with one resonance for
$\mathbb{J}_{+}=1$, $\mathbb{I}_{+}={\mathbb{J}_{+}}/{2}$,
and  for the fixed point $(N_1,\theta_1) = ({\mathbb{J}_{+}}/{4},2.8)$}
\end{figure}

\begin{figure}
\caption{\label{Figure10} Plots of $\eta(x,t)$ for  Hamiltonian (\ref{504})
 with two resonances
 and for different values of $\alpha$:
      $(a)$ standing wave $\alpha=1$, $(b)$ stationary wave $\alpha$ indefinite $c=0$,
    $(c)$ traveling wave $\alpha=-1$, $(d)$ time periodic wave $\alpha \in \mathbb{Q}$
    ($-\mathbb{J}_{+}+2\mathbb{J}_{-} =J_0$),
    $(e)$ time quasiperiodic wave $\alpha =\sqrt{0.4}$.}

\caption{\label{Figure11}
Poincar\'e sections for  Hamiltonian (\ref{HamAA}) with two resonances
for $H=2$,  $ \mathbb{J}_{+}=\mathbb{J}_{-}=1$
 and for different values of $ \mathbb{I}_{+}:$
  $ (a) \mathbb{I}_{+}=\mathbb{I}_{-}=0.3$,  $ (b) \mathbb{I}_{+}=\mathbb{I}_{-}=0.4$,
  $ (c) \mathbb{I}_{+}=\mathbb{I}_{-}=0.5$,
  $ (d) \mathbb{I}_{+}=\mathbb{I}_{-}=0.8$. The grey region is the
  accessible region where
 $\rm{max}(0,\mathbb{K}_{+}) \leq N_1 \leq {\mathbb{I}_{+}}/{2}.$}

\caption{\label{Figure12}
Time evolution of $N_{j}$ for $j=-1,-2,-3,1,2,3$,
for  Hamiltonian (\ref{HamAA}) with two resonances
and  for $\mathbb{I}_{+}=\mathbb{I}_{-}=0.5$ and
for an initial condition in the chaotic region $(N_1,\theta_1)=(0.152,-3.07).$ }

\caption{\label{Figure13}
Shape of the wave  $\eta(x,t)$ for $\mathbb{I}_{+}=\mathbb{I}_{-}=0.5$
for  Hamiltonian (\ref{HamAA}) with two resonances
and for an initial condition in the chaotic region $(N_1,\theta_1)=(0.152,-3.07).$}
\end{figure}

\end{document}